\documentclass[apj]{emulateapj}

\usepackage{amsmath}
\usepackage{amssymb}
\usepackage{natbib}
\usepackage{graphicx}
\usepackage{multirow}
\usepackage{booktabs}
\usepackage{bm}

\newcommand{\acsb}{$B_{435}$}

\newcommand{\acsv}{$V_{606}$}
\newcommand{\acsi}{$i_{775}$}
\newcommand{\acsz}{$z_{850}$}

\newcommand{\sersic}{S\'ersic}

\newcommand{\zspec}{$z_{\mathrm{spec}}$}
\newcommand{\zphot}{$z_{\mathrm{phot}}$}
\newcommand{\params}{(\alpha, M^*, R_0, \sigma_{\ln R_e}, \beta)}
\newcommand{\fullparams}{(\alpha, M^*, \phi^*, R_0, \sigma_{\ln R}, \beta)}

\newcommand{\GF}{\mathrm{GF}}
\newcommand{\alpharesb}{-1.68}
\newcommand{\alpharesberr}{^{+0.068}_{-0.095}}
\newcommand{\mstarresb}{-20.60}
\newcommand{\mstarresberr}{^{+0.13}_{-0.17}}
\newcommand{\phistarresb}{1.79}
\newcommand{\phistarresberr}{^{+0.32}_{-0.52}}
\newcommand{\Rzeroresbas}{0.19}
\newcommand{\Rzeroresbaserr}{^{+0.014}_{-0.016}}
\newcommand{\Rzeroresbkpc}{1.34}
\newcommand{\Rzeroresbkpcerr}{^{+0.099}_{-0.108}}
\newcommand{\sigmaresb}{0.83}
\newcommand{\sigmaresberr}{^{+0.046}_{-0.044}}
\newcommand{\betaresb}{0.22}
\newcommand{\betaresberr}{^{+0.058}_{-0.056}}
\newcommand{\alpharesv}{-1.74}
\newcommand{\alpharesverr}{^{+0.15}_{-0.20}}
\newcommand{\mstarresv}{-20.53}
\newcommand{\mstarresverr}{^{+0.24}_{-0.27}}
\newcommand{\phistarresv}{1.55}
\newcommand{\phistarresverr}{^{+0.62}_{-0.77}}
\newcommand{\Rzeroresvas}{0.19}
\newcommand{\Rzeroresvaserr}{^{+0.025}_{-0.033}}
\newcommand{\Rzeroresvkpc}{1.19}
\newcommand{\Rzeroresvkpcerr}{^{+0.21}_{-0.16}}
\newcommand{\sigmaresv}{0.90}
\newcommand{\sigmaresverr}{^{+0.15}_{-0.065}}
\newcommand{\betaresv}{0.25}
\newcommand{\betaresverr}{^{+0.15}_{-0.14}}

\begin{document}

\title{The Bivariate Size-Luminosity Relations for Lyman Break Galaxies\\at $z\sim 4-5$}

\author{Kuang-Han Huang}
\affil{Johns Hopkins University, 3400 N. Charles Street, Baltimore, MD 21218 USA; kuanghan@pha.jhu.edu}

\author{Henry C. Ferguson}
\affil{Space Telescope Science Institute, 3700 San Martin Drive, Baltimore, MD 21218}

\author{Swara Ravindranath}
\affil{The Inter-University Center for Astronomy and Astrophysics, Pune University Campus, Pune 411007, Maharashtra, India}

\and\author{Jian Su}
\affil{Johns Hopkins University, 3400 N. Charles Street, Baltimore, MD 21218 USA}
\affil{Department of Radiology, Mayo Clinic, Rochester, MN 55905}

\begin{abstract}
We study the bivariate size-luminosity distribution of Lyman Break Galaxies (LBGs) selected at redshifts around 4 and 5 in GOODS and the HUDF fields. We model the size-luminosity distribution as a combination of log-normal distribution (in size) and Schechter function (in luminosity), therefore it enables a more detailed study of the selection effects. We perform extensive simulations to quantify the dropout-selection completenesses and measurement biases and uncertainties in two-dimensional size and magnitude bins, and transform the theoretical size-luminosity distribution to the expected distribution for the observed data. Using  maximum-likelihood estimator (MLE), we find that the Schechter function parameters for \acsb-dropouts are $\alpha=\alpharesb\alpharesberr$, $M^*=\mstarresb\mstarresberr$, and $\phi^*=\phistarresb\phistarresberr\times 10^{-3}$ $\mathrm{Mpc}^{-3}$. The log-normal size distribution is characterized by the peak $R_0=\Rzeroresbkpc\Rzeroresbkpcerr$ kpc at $M_{1500}=-21$ mag, width $\sigma_{\ln\mathrm{R}}=\sigmaresb\sigmaresberr$, and the slope of the size-luminosity ($RL$) relation $\beta=\betaresb\betaresberr$. Similarly, for \acsv-dropouts we find $\alpha=\alpharesv\alpharesverr$, $M^*=\mstarresv\mstarresverr$, $\phi^*=\phistarresv\phistarresverr\times 10^{-3}$ $\mathrm{Mpc}^{-3}$, $R_0=\Rzeroresvkpc\Rzeroresvkpcerr$ kpc, $\sigma_{\ln\mathrm{R}}=\sigmaresv\sigmaresverr$, and $\beta=\betaresv\betaresverr$. The Schechter function parameters are consistent with the values in the literature, while the size distributions are wider than expected from the angular momentum distribution of the underlying dark matter halos. The slope of the size-luminosity ($RL$) relation is similar to those found for local disk galaxies, but considerably shallower than local early-type galaxies.
 
\end{abstract}
\keywords{galaxies: evolution --- galaxies: high-redshift --- galaxies: luminosity function --- galaxies: structure --- method: data analysis}

\section{Introduction}\label{sec:intro}
The study of the high-redshift galaxies has benefited tremendously from the launch of the \textit{Hubble Space Telescope} (\textit{HST}), especially from its exquisite sensitivity and angular resolution. In particular, the Advanced Cameras for Surveys (ACS) on board \textit{HST} observes in the optical wavelength range, probing the rest-frame far-UV wavelengths of the emission from high-redshift galaxies that are sensitive to the instantaneous star-formation rate. Several deep surveys of field galaxies, such as the Great Observatories Origins Deep Survey (GOODS; \citealt{Giavalisco.2004a}) and the Hubble Ultra Deep Field Survey (HUDF; \citealt{Beckwith.2006}), took advantage of the powerful survey abilities of ACS to study the high-redshift universe. With multi-wavelength data taken by other instruments, the GOODS fields are among the most data-rich portions of the sky, looking back to the very early universe.

The Lyman Break selection method, pioneered by \cite{Steidel.1992}, selects high-redshift galaxies using their rest-frame UV continuum emission.  The Lyman Break technique selects star-forming galaxies with redshift between 3 and 6 using near-UV and optical photometry (and higher-redshift galaxies using near-IR data). The Lyman Break Galaxies (LBGs) are in general actively star-forming, UV-bright galaxies that are possibly the progenitors of present-day massive spheroidal galaxies (e.g. \citealt{Pettini.2001,Shapley.2001,Papovich.2001}). Their rest-frame UV properties shed light on the cosmic history of star formation (\citealt{Madau.1996,Giavalisco.2004b}), although the attenuation of UV light by dust in these galaxies also plays an important role. LBGs are an important population of galaxies in the early universe, when the cosmic star-formation rate density was much higher than at $z \sim 0$.

The luminosity function, i.e. the number-density distribution as a function of galaxy luminosity, is one of the most important and the most extensively studied statistical properties of the high-redshift galaxy population. Specifically, at redshift around 4 and 5, many studies of the UV luminosity function (usually parameterized by the Schechter function; \citealt{Schechter.1976}) of the LBGs already exist (\citealt{Steidel.1999,Ouchi.2004,Giavalisco.2005,Yoshida.2006,Beckwith.2006,Iwata.2007,Bouwens.2007,Oesch.2007,vanderBurg.2010}). However, differences in sample selection, photometry, and modeling procedures can lead to contradictory results using the same datasets. For example, \cite{Beckwith.2006} derived UV luminosity functions of LBGs at $z \sim$ 4, 5, and 6 selected from the HUDF and the GOODS fields. They concluded that the Schechter function parameters $\alpha$ and $M^*$ do not evolve significantly between these redshifts, but the overall normalization constant $\phi^*$ evolves by a factor of $\sim 3$ in the redshift range $4 < z < 6$. On the other hand, the analyses of \cite{Bouwens.2007} show that the Schechter function parameters $M^*$ brightens by $\sim$ 0.7 mag, but $\phi^*$ does not seem to evolve much. The discrepancies remind us of the complexities involved in deriving the luminosity functions. 

The size distribution (or equivalently the surface-brightness distribution) of LBGs is not modeled jointly with luminosity function in previous studies. A variety of approaches have been adopted to account for the bias against selecting low-surface-brightness galaxies and biases in their photometry. Perhaps the most sophisticated approach was that of \citet{Bouwens.2007}, which uses a combination of "cloning" lower-redshift dropout galaxies to higher redshift, and photometrically degrading galaxies observed in deep observations to simulate shallower observations (see also \citealt{Oesch.2007} and their ``dimmed galaxies'' simulations). The appeal of this approach is that it uses real galaxies, with their sometimes clumpy and irregular morphologies. However, there are some drawbacks to the approach that suggest at least checking the results with a different approach altogether. Among the drawbacks are the following: (1) Only relatively small numbers of galaxies with high S/N are available for cloning to higher redshift.  (2) At $z >1.2$, angular diameter increases with redshift, which implies that the cloned images should be made bigger. Evolution toward smaller physical sizes at higher redshifts  may be more than sufficient to counteract this effect. But even in that case, the formally correct procedure would involve deconvolving, rescaling, and re-convolving (or equivalently convolving with an appropriate kernel that accomplishes this).  In practice, this is probably a minor effect and it has been generally ignored.  (3) The true physical sizes of the cloned galaxies are not really known; all that is known is their measured sizes prior to cloning. (4) When a deep field is used to provide sources for assessing completeness and photometric bias in a wider field, this is of course attempting to represent a large volume as a bootstrap resampling of a small volume.  Rare objects are likely to be poorly represented in the small volume, and thus the completeness and photometric bias in those parts of parameter space are likely to be poorly characterized for the wide images.  (5) These corrections make assumptions about the size distribution and its evolution. These become strong priors on the fit to the luminosity function parameters, leaving one somewhat uneasy about the robustness of the results.

An alternative approach, which we adopt here, is to model the luminosity function jointly with either the size or the surface-brightness distribution to include the surface-brightness-related selection effects. Ideally, one would also like to include other physical parameters that influence sample selection jointly such as color, Ly$\alpha$ equivalent width and surface-brightness profile. But the problem quickly becomes intractable as the number of dimension increases. In this paper we attempt to model the size distribution jointly with the luminosity function in order to remove some of the biases in the UV luminosity function. An important motivation for fitting the bivariate distribution is to reveal the degeneracy between the luminosity-function parameters and the size-distribution parameters. Because the observations select strongly against the low-surface-brightness galaxies, it is not obvious without detailed study that the LF parameters are robust. We do not simultaneously fit for the distributions of color, Ly$\alpha$, or surface brightness profile here --- in part because this raises the dimensionality of the problem enough that exploring parameter space becomes difficult, and in part because the selection of functional forms for those distributions is less obvious. So instead of fitting those parameters, we follow the standard practice of verifying that our model distributions qualitatively match the observed distributions, and focus on exploring the effect of including size in the luminosity function modeling.

Galaxy size also provides important constraints on the galaxy-formation models. The canonical picture of disk formation introduced by \cite{Fall.Efstathiou.1980} (see also \citealt{Fall.1983,Dalcanton.1997,Mo.1998}) states that disks form from the gas within dark matter (DM) halos collapsing while approximately conserving its angular momentum; the angular momentum comes from the tidal interactions between dark matter halos around the trans-linear regime in the hierarchical structure formation picture. The angular momentum of DM halos is usually parameterized by the spin parameter $\lambda\equiv J|E|^{1/2}G^{-1}M^{-5/2}$ (\citealt{Peebles.1969}), and it follows log-normal distribution with peak $\bar{\lambda}\sim 0.05$ and $\sigma_{\ln\lambda}\sim0.5$ in $N$-body cosmological simulations, relatively independent of halo mass or redshift (\citealt{Barnes.1987,Warren.1992,Bullock.2001}). Assuming that the baryons and dark matter receive the same amount of angular momentum per unit mass by tidal torque, one can relate the halo virial radius to the size of the disk: $R_\mathrm{d} \propto \lambda\,R_{\mathrm{vir}}$. Therefore the size distribution should reflect the underlying $\lambda$ distribution, because gas acquired most of its angular momentum with DM halos before collapse. On the other hand, reasonable galaxy sizes have been notoriously difficult to reproduce in the hydrodynamical simulations. Early investigations encountered the ``angular-momentum problem'' (e.g. \citealt{Navarro.1991,Steinmetz.1999}) that lead to unrealistically small galaxy sizes in the simulations. Later investigations implemented various forms of feedback to alleviate the angular-momentum problem (e.g. \citealt{Sales.2010,Brooks.2011}), and improvements in numerical treatments of hydrodynamics have also shown promises in eliminating the angular momentum problem without resorting to late feedback (e.g. \citealt{Keres.2011,Torrey.2011}). 

The rest-frame UV size distribution of LBGs was also studied by several authors (independent of the luminosity function). For example, \citet{Ferguson.2004} studied the mean half-light radii of star-forming galaxies (including LBGs) in the GOODS fields from $z \sim 1$ to $z \sim 5$. Also \citet{Hathi.2008} studied the sizes of the spectroscopically-confirmed LBGs between $z \sim 3 - 6$. Both papers found that the evolution of the mean half-light radii roughly follows the $H^{-1}(z)$ relation, where in the concordance cosmology $H^2(z) =\Omega_m(1+z)^3+\Omega_\Lambda$. \citet{Bouwens.2004} and \citet{Oesch.2010} studied the mean size of LBGs from $z \sim 2.5$ to $z \sim 8$ and found that the evolution of the mean half-light radii of LBGs within the same luminosity range follow the $(1+z)^{-1}$ relation, roughly corresponding to $H^{-2/3}(z)$ at $z\gtrsim 2.5$. Building upon previous works, we investigate the size distribution of LBGs by fitting the size distribution jointly with luminosity function and exploring the degeneracy between the two distributions, incorporating measurement biases calibrated to known half-light radii of parametric galaxy profiles. The cloning technique can be used to create realistic galaxy profiles, but the uncertainties in the true sizes of the cloned galaxies make it difficult to calibrate our corrections. We also constrain the scatter around the mean size-luminosity relation, which is an additional constraint on the processes that govern star formation and feedback at early times. 

Because surface brightness affects both the detectability of high-$z$ galaxies and the accuracy of the measurement of magnitude and size, we model the systematic biases in the measurements of both magnitude and size, which has not been done before on high-$z$ LBGs. There were attempts to derive either the size-luminosity or the size-stellar mass distributions of galaxies at $z<3$ (e.g. \citealt{Simard.1999,deJong.2000,Trujillo.2004,Trujillo.2006,Mosleh.2011}), and after the first submission of this paper, simultaneous and independent efforts by \citet{Mosleh:2012cw} and \citet{Grazian:2012uh} also investigated these distributions at $z>4$, but not with the technique used here.

\textit{Structure of the paper.} --- In this paper we study the bivariate size-luminosity distribution of LBGs selected at redshifts around 4 and 5. In Section \ref{sec:sample} we describe our sample selection with optical color criteria. We also describe how we estimate the ``photometric interloper'' fractions --- fraction of our sample that are low-$z$ sources scattered into the color-selection window due to photometric scatter --- statistically, and compare that with the available spectroscopic redshifts in GOODS. In Section \ref{sec:measure} we describe how we measure the galaxy magnitude and size, and why we opt for using GALFIT (\citealt{Peng.2010}) instead of SExtractor (\citealt{SExtractor}) for size measurement. In Section \ref{sec:modeling} we describe the size-luminosity distribution model we adopt, and the simulations we perform in order to derive selection completenesses, measurement biases, and measurement uncertainties \textit{in magnitude and size bins}. We then explain how we use the derived kernels to transform the model to produce the expected distribution of observations, and how we perform the fit to derive the best-fit parameters. In Section \ref{sec:bestfitpar} we present the results of the fitting together with the confidence intervals of each parameters. In Section \ref{sec:discussion} we discuss some implications of our results, including comparison with previous LF determinations, size evolution, width of the size distribution, and the size-luminosity relation. We summarize our work in Section \ref{sec:summary}. We adopt the cosmological parameters $H_0=70\,\mathrm{km}\,\mathrm{s}^{-1}\,\mathrm{Mpc}^{-1}$, $\Omega_m=0.3$, and $\Omega_\Lambda=0.7$ for easy comparisons with previous results, and these values are very close to the WMAP7+BAO+$H_0$ mean cosmological parameters (\citealt{Komatsu.2011}) of $H_0=70.2\,\mathrm{km}\,\mathrm{s}^{-1}\,\mathrm{Mpc}^{-1}$, $\Omega_m = 0.275$, and $\Omega_\Lambda = 0.725$. The magnitudes are in the AB magnitude system (\citealt{Oke.1983}).

\section{Lyman Break Galaxy Sample Selection}\label{sec:sample}
\indent We select Lyman Break Galaxies using colors computed from the public GOODS ACS version 2 (v2) catalogs, excluding the HUDF region. For more details of how the catalogs are prepared please see \citet{Giavalisco.2004a} and the documentation on the GOODS website\footnote{\url{http://archive.stsci.edu/prepds/goods/}}. In order to constrain the statistics at the faint end of the luminosity function, we also select LBGs from the Hubble Ultra Deep Field (HUDF; \citealt{Beckwith.2006}) dataset. HUDF dataset consists of ACS images in the same passbands used in the GOODS ACS dataset (\acsb, \acsv, \acsi, and \acsz) covering an 11 arcmin$^2$ area within the GOODS-S field. We create our own catalogs in the HUDF based on detection in \acsz-band and photometry on all bands in dual-image mode of SExtractor, using 0.5 for \texttt{DETECT\_THRESH} and \texttt{ANALYSIS\_THRESH}, and 0.03 for \texttt{DEBLEND\_MINCONT}. For details on the ACS imaging product in the HUDF please see \cite{Beckwith.2006}.

\subsection{Selection Criteria}\label{subsec:selcrit}

LBG color-selection criteria are designed for selecting
star-forming galaxies at high redshift using two features in the
spectral energy distributions (SED) of star forming galaxies
(e.g. \citealt{Steidel.1992,Steidel.1996,Giavalisco.2002}). One
feature is the spectral break at rest-frame 912 \AA\ due to the
absorption of UV photons by the stellar atmosphere of early-type stars
and the neutral hydrogen in the interstellar medium (ISM) surrounding
the star-forming regions. The second feature is the blanketing of 
Lyman-series lines by the intervening neutral hydrogen in the intergalactic
medium (IGM) along the lines of sight to
galaxies (\citealt{Madau.1995,Meiksin.2006}). Therefore it is possible to
use the colors between a set of broadband filters to target a specific
redshift range.

We adopt the color-selection criteria in 
\citet{Giavalisco.2004b}, which are optimized to select LBGs at $z \sim 4$ and $z \sim 5$.
The criteria for the \acsb-dropouts are

\begin{equation}\label{eq:bdrops}
   \begin{aligned}
      (B_{435} - V_{606}) & \geq 1.2 + 1.4 \ (V_{606} - z_{850})\ \wedge \\
      (B_{435} - V_{606}) & \geq 1.2\ \wedge\ (V_{606} - z_{850}) \leq 1.2
   \end{aligned}
\end{equation}

and the criteria for the \acsv-dropouts are

\begin{equation}\label{eq:vdrops}
   \begin{aligned}
      & \{[(V_{606}-i_{775}) > 1.5 + 0.9 \ (i_{775}-z_{850})]\ \vee\\
      & [(V_{606}-i_{775}) > 2.0]\}\ \wedge\ (V_{606}-i_{775}) \geq 1.2\ \wedge \\
      & (i_{775}-z_{850}) \leq 1.3,
   \end{aligned}
\end{equation}

where $\wedge$ is the logical ``AND" operator and $\vee$ is the logical ``OR" operator. 
The colors are computed using 
the isophotal magnitudes \texttt{MAG\_ISO} reported by SExtractor in order to
obtain consistent colors. All magnitudes quoted afterwards are the \texttt{MAG\_AUTO} from SExtractor 
unless otherwise stated.

We select galaxies that satisfy the above color criteria down to \acsz$=26.5$ mag in GOODS and \acsz$=28.5$ mag in HUDF, while also demanding that the signal-to-noise ratio ($S/N$) in the \acsz-band be greater than 5. For \acsv-dropouts we also require all sources to have \acsb-band $(S/N)_B \leq 5$, as at $z \sim 5$ the \acsb-band is entirely blue-ward of the Lyman break and should not have significant detection. This $S/N$ criterion is more liberal than those in the literature (e.g. \citealt{Giavalisco.2004b}). The reason for adopting a liberal cut here is to increase the completeness of our sample, as our simulations show a non-negligible fraction of the true $z\sim 5$ sources whose $S/N$ in \acsb-band are slightly higher than 2 (the value commonly used in the literature) because the SExtractor-estimated $S/N$ becomes unreliable for the non-detected sources. Among the 52 sources with $2 < S/N < 5$ in \acsb-band, 6 are broken-up diffraction spikes, 46 have $S/N < 3$, 8 have $3 < S/N < 4$, and only 2 have $4 < S/N < 5$. Furthermore, all three of the sources with spectroscopic redshifts that have $S/N > 2$ in \acsb-band are at $z_{\mathrm{spec}}>4.4$ and not low-$z$ interlopers. Because the choice in the $S/N$ limit is photometry dependent and has some intrinsic uncertainty, we decide to use a more liberal threshold in the \acsb-band $S/N$ to increase our completeness without introducing a large number of interlopers. We also exclude sources with SExtractor-reported \texttt{CLASS\_STAR} $\geq 0.95$ down to \acsz$\leq 26$ mag in GOODS and \acsz$\leq 28$ mag in the HUDF to eliminate field stars. The resultant sample still contains two AGN candidates (IAU ID J033216.86-275043.9 and J033243.84-274918.1; \citealt{Wolf.2004}); we therefore exclude these two sources from further analyses. We next visually inspect each selected dropout source to remove other image artifacts that are not removed in the image-reduction process, and this removes 16 and 19 sources from the \acsb-dropout and \acsv-dropout sample, respectively, all of which are broken diffraction spikes from bright stars. 

At the bright end of our samples (\acsz$\leq24.5$ mag in GOODS and \acsz$\leq 25.5$ mag in the HUDF) the majority of sources have spectroscopic redshifts. For the few sources that do not, all of them have photometric redshifts from \citet{Dahlen.2010} (we describe the redshift information of our sample in Section \ref{subsec:zdist}). Therefore we eliminate the bright sources that are confirmed low-$z$ interlopers (defined as $z<3$ for the \acsb-dropout sample and $z<4$ for the \acsv-dropout sample) based on available redshift information. Doing so removes 1 \acsb-dropout and 3 \acsv-dropouts from GOODS. In fainter magnitude bins (\acsz$>24.5$ mag in GOODS and \acsz$>25.5$ mag in the HUDF) we do not have redshift information for most sources, so we will model the interloper fractions in Section \ref{subsec:interloper} and include them as an additional component of our model.

In addition, if a given source exists in both the GOODS and HUDF source lists, we use the values measured in the HUDF because it is deeper. With the above criteria we have a total of 1063 \acsb-dropouts and 465 \acsv-dropouts. The number of dropouts in the two GOODS fields (excluding the deeper HUDF), 384 \acsb-dropouts and 196 \acsv-dropouts in GOODS-N and 454 \acsb-dropouts and 157 \acsv-dropouts in GOODS-S, are consistent within the cosmic variance estimated by the cosmic variance calculator of \cite{Trenti.2008}, which is about $15\%$. The numbers are summarized in Table \ref{tab:dropouts}.

\begin{deluxetable*}{ccccccc}
   \tablecaption{The number of selected dropouts. \label{tab:dropouts}}
   \tablehead{
      \colhead{Dropout} & \colhead{Field} & \colhead{\acsz\ limit\tablenotemark{a}} & \colhead{Number} 
      & \colhead{Number\tablenotemark{b}} & \colhead{Number\tablenotemark{c}}\\
      \colhead{} & \colhead{} & \colhead{(mag)} & \colhead{(SExtractor sample)} \ &\colhead{(GALFIT sample)}
      & \colhead{(GALFIT quality sample)}}
   \startdata 
   \multirow{2}{*}{$B$-dropouts}  &  GOODS  &  $26.5$  &  $838$ & $733$ & $586$ \\
     &  HUDF   &  $28.5$  &  $225$ &  $202$  &  $146$ \\
   \midrule
   \multirow{2}{*}{$V$-dropouts}  &  GOODS  &  $26.5$  &  $353$ & $325$ &  $219$ \\
     &  HUDF   &  $28.5$  &  $112$  &  $96$  &  $61$ 
   \enddata
\tablenotetext{a}{\acsz-band \texttt{MAG\_AUTO} limit for sample selection.}
\tablenotetext{b}{Number of sources with GALFIT-returned magnitudes (in \acsi-band for \acsb-dropouts and in \acsz-band for \acsv-dropouts) brighter than 26.5 mag in GOODS and 28.5 mag in HUDF.}
\tablenotetext{c}{Number of sources that have acceptable GALFIT fits.}
\end{deluxetable*}

\subsection{Redshift Distributions}\label{subsec:zdist} 
\begin{figure}[t]
	\epsscale{1.3}
   \plotone{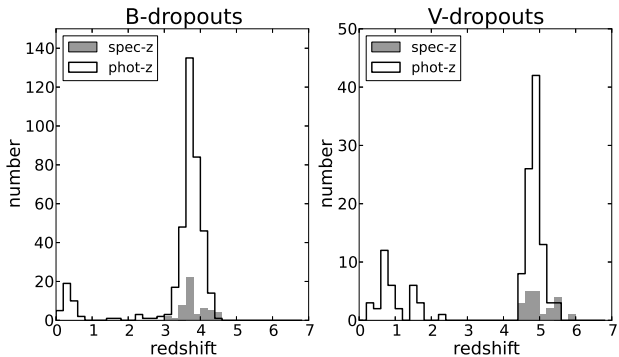}
   \figcaption{Redshift distribution of \acsb-dropouts and
\acsv-dropouts selected from GOODS-S. The bin width of the redshift axis
is $\Delta z = 0.2$. Photometric redshifts are from
\citet{Dahlen.2010}. References for the spectroscopic redshift data are
given in Section \ref{subsec:zdist}.\label{fig:zhist}}
\end{figure}

There have been several campaigns to measure spectroscopic redshifts (\zspec) in the GOODS fields that cover the sources of interest here (\citealt{Steidel.1999,Cristiani.2000,Cohen.2000,Cohen.2001,Steidel.2003,Wirth.2004,Daddi.2009a,Daddi.2009b,Barger.2008,Vanzella.2009,Popesso.2009}; Stern et al. 2012, in preparation). In addition, \citet{Dahlen.2010} produced the deepest photometric-redshift (\zphot) catalog in GOODS-S (the photometric-redshift catalog of \citealt{Cardamone.2010} in GOODS-S, although derived with more bands, does not go as deep as \citealt{Dahlen.2010}.) In Figure \ref{fig:zhist} we show the redshift distribution of our selected sources (we only show spectroscopic redshifts with good or acceptable qualities if such flags exists in the catalogs.) In general the color-selection criteria are able to select \acsb-dropouts within the redshift range $3.4 \lesssim z \lesssim 4.4$ and \acsv-dropouts within $4.4 \lesssim z \lesssim 5.6$. We do not use the redshift information to identify interlopers and exclude them from analysis in the faint end, because the spectroscopic redshifts are incomplete and the photometric redshifts have larger errors. Instead we try to model the interloper fractions in Section \ref{subsec:interloper} and include their contribution as an additional component in our model.

High-redshift, star-forming galaxies have a higher fraction of Ly$\alpha$-emitters (LAEs) than local star-forming galaxies (see e.g. \citealt{Reddy.2008,Finkelstein.2009,Cowie.2010,Ciardullo.2012}). The fraction of star-forming galaxies at $z \gtrsim 5$ with large rest-frame Ly$\alpha$ equivalent widths could also be a function of redshift and luminosity (see e.g. \citealt{Stark.2010,Schaerer.2011}). At $z\sim 5.6 - 5.8$, Ly$\alpha$ is redshifted into the \acsi-band, therefore a galaxy with strong Ly$\alpha$ emission (rest-frame equivalent width $W_{\mathrm{Ly\alpha}}\gtrsim 50$ \AA) will have redder \acsv$-$\acsi\ and bluer \acsi-\acsz\ colors than the colors determined by the continuum alone, making them more likely to be selected as \acsv-dropouts instead of \acsi-dropouts (see e.g. \citealt{Malhotra.2005,Dow-Hygelund.2007,Stanway.2007,Stanway.2008,Bouwens.2007,Stark.2010}). Among our \acsv-dropouts with spectroscopic redshifts, 6 source out of 40 (15\%) have $z_{\mathrm{spec}}\geq 5.5$, representing a secondary peak of $z_{\mathrm{spec}}$ distribution. It is possible that these sources have strong Ly$\alpha$-emission-perturbed broadband colors. However, at $z\geq 5.5$ there is yet to be a complete census of Ly$\alpha$ equivalent widths, so we can not model the effects of Ly$\alpha$ emission on our sample selection accurately. A simple estimate of the possible fraction of \acsv-dropouts that are $z\sim 6$ galaxies with strong Ly$\alpha$ emission --- using the best-fit exponential distribution of Ly$\alpha$ equivalent widths of LAEs at $z\sim 3.1$ (\citealt{Ciardullo.2012}) and the number densities of \acsi-dropouts from \citet{Bouwens.2007} --- suggests that these galaxies make up at most about 5\% of our \acsv-dropout sample in GOODS. Therefore we do not consider $z\sim 6$ star-forming galaxies with strong Ly$\alpha$ emission to contribute significantly to the \acsv-dropout sample.

\subsection{Interloper Fractions from Photometric Scatter}\label{subsec:interloper}
Sources other than high-$z$ star-forming galaxies could also be selected by the LBG criteria. One type of interlopers are sources with intrinsically red SEDs such as dwarf stars (of spectral types M, L, and T) and $z\leq 2$ quiescent galaxies (\citealt{Ryan.2005,Pirzkal.2005, Pirzkal.2009,Steidel.1999,Yoshida.2006}). Another type of interlopers are sources with intrinsic SEDs outside of the target range, but with colors close enough to the selection boundaries that they are scattered into the color-selection window due to photometric error. To minimize the first type of interlopers, we identify low-$z$ interlopers with available redshift information in the bright end in Section \ref{subsec:selcrit}. At the faint end, identifying individual interlopers of either type becomes intractable due to the limited spectroscopic information. In this section, we estimate the interloper contributions from photometric scatter (hereafter ``photometric interlopers''). 

We estimate the photometric interloper fractions at \acsz$>24.5$ mag in GOODS and \acsz$>25.5$ mag in the HUDF, in magnitude bins of width 0.5 mag, with a set of simulations (the ``photometric interloper simulations''). Our simulations are very similar to the ones performed in \citet{Bouwens.2006} (where they estimated the contamination fraction in the HUDF) and \citet{Iwata.2007}. In the photometric interloper simulations we use all sources with $S/N \geq 5$ in \acsz-band and $S/N \geq 3$ in \acsb-, \acsv-, and \acsi-bands that are not selected as dropouts. Only sources with either spectroscopic redshift or photometric redshift much lower than $z=3$ are used in order to minimize the contribution of sources that are just below the selection redshift range. The sources selected this way have reasonably constrained SED shapes that span a wide range. For each selected source we then rescale its \acsz-band magnitude to a randomly selected value between 24.5 and 26.5 mag (in GOODS) and between 25.5 and 28.5 mag (in HUDF). After rescaling the magnitudes in other bands by the same factor, we randomly perturb its magnitude in each band according to the appropriate photometric error in the given band and magnitude bin (assuming Gaussian error in magnitude). We calculate the number of these sources that fall within the color-selection window in each magnitude bin, and also calculate their fractional contribution among the real dropouts, defined as 
\[ f_{\mathrm{interloper}} = \frac{\mathrm{num.\ of\ expected\ photometric\ interlopers}}
{\mathrm{num.\ of\ selected\ dropouts}}\]
The same procedure is repeated 5000 times to collect good statistics. 

\begin{figure}[t]
   \plotone{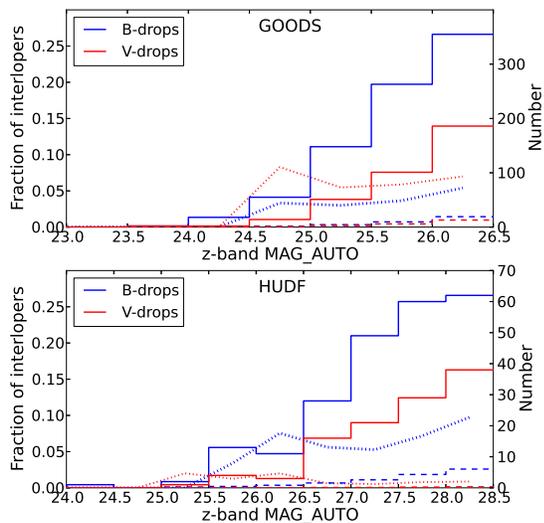}
   \figcaption{The estimated interloper fractions due to photometric scatter as a function of the \acsz-band magnitude. The top panel shows the results in GOODS-depth images; the bottom panel shows the result in the HUDF-depth images. Black histograms and lines are for \acsb-dropouts; gray histograms and lines are for \acsv-dropouts. Solid histograms are the observed number of dropouts; dashed histograms are the expected number of interlopers; dotted lines are the estimated interloper fraction in each magnitude bin. In most bins the interloper fractions are below 10\%; these fractions are likely upper limits because sources that are scattered into the color-selection window could still be rejected using other criteria that cull genuine high-$z$ galaxies. The numbers (of sources in each bin) are read off from the right ordinate, and the interloper fractions are read off from the left ordinate. In bins where no real dropouts are selected, the interloper fractions are set to zero, and no additional number densities from photometric interlopers are added to the model. 
\label{fig:contam_frac}}
\end{figure}

The results of the photometric interloper simulations are presented in Figure \ref{fig:contam_frac}. The estimated overall photometric interloper fractions are below $8\%$, and for the \acsv-dropouts in the HUDF the photometric interloper fraction is especially low, possibly owing to the extremely deep \acsi-band data in HUDF that constrain the \acsi-\acsz\ color well.  In our photometric-interloper simulations, we do not reject sources using the \acsb-band $S/N$ threshold for \acsv-dropouts, so the true photometric interloper fractions are likely even lower. In general, the photometric-interloper fractions stay relatively constant over all magnitude bins probed.

\begin{deluxetable}{ccrrr}
   \tablecaption{The photometric interloper fractions at \acsz$>24.5$ mag in GOODS and \acsz$>25.5$ mag in the HUDF from the photometric interloper simulations \label{tab:interlopers}}
   \tablehead{\colhead{Dropout}&\colhead{Field}&\colhead{$N_{\mathrm{interloper}}$}
   &\colhead{$N_{\mathrm{dropout}}$}&\colhead{$f_{\mathrm{interloper}}$}}
   \startdata
   \acsb-dropouts & GOODS & 35.1 & 821 & $4.3\%$ \\
   \acsb-dropouts & HUDF   & 15.8 & 223 & $7.1\%$ \\
   \acsv-dropouts & GOODS & 22.9 & 352 & $6.5\%$ \\
   \acsv-dropouts & HUDF   & 0.9  & 111  & $0.8\%$ 
   \enddata
   \tablecomments{$N_{\mathrm{interloper}}$ is the expected number of photometric interlopers
   from simulations, while $N_{\mathrm{dropout}}$ is the number of real LBGs
   selected in the given magnitude range.}
\end{deluxetable}

As verification of the photometric interloper fractions from simulations, we compare the estimated photometric interloper fractions $f_{\mathrm{interloper}}$ with those estimated from available spectroscopic redshifts ($z_{\mathrm{spec}}$) alone. Among the \acsb-dropout sample in GOODS, 90 sources have secure spectroscopic redshifts at \acsz$>24.5$ mag, 1 of them have $z_{\mathrm{spec}}<3$ ($1.1\%$). Similarly, among the \acsv-dropout sample in GOODS, 37 sources have secure spectroscopic redshifts at \acsz$>24.5$ mag, and none of them have $z_{\mathrm{spec}}<4$. In the HUDF, only 4 \acsb-dropouts and 4 \acsv-dropouts with \acsz$>25.5$ mag have $z_{\mathrm{spec}}$, and none of them are low-$z$ interlopers. All of the interloper fractions estimated from available spectroscopic redshifts are consistent with what we find from the photometric interloper simulations. In all cases, the estimated interloper fractions from the simulations are higher than those estimated from the sources with spectroscopic redshifts. Because this set of simulations accounts for objects scattered into the color-selection window, but does not use other criteria such as stellarity or image flags to reject sources, the interloper fractions estimated are likely upper limits.

\section{Magnitude and Size Measurement}\label{sec:measure}
Defining and measuring the size of a galaxy from the image is a long-standing problem. In the past there have been several definitions of galaxy size, all with the general aim of being unbiased and easy-to-measure. For example, the Petrosian radius (\citealt{Petrosian.1976,Graham.2005} and references therein) is defined through the Petrosian index $\eta(R) = \langle I \rangle_R / I(R)$, the average surface brightness within radius R over the surface brightness at R. Petrosian radius is then defined as a multiple of $R_P$ at which $\eta(R_P)$ (or $1/\eta(R_P)$) equals some chosen value. In practice, one cannot measure $I(R)$, but instead one measures $\langle I(R_1, R_2)\rangle$, the mean surface brightness in some range of radius, which is subject to resolution effects at small $R$ and $S/N$ limitations at large $R$. Another common definition is the half-light radius (e.g. \citealt{Bershady.2000}) that is the radius within which half of the total light from the galaxy is enclosed. However it is sensitive to the aperture adopted to measure the total light from the galaxy. On the other hand, the effective half-light radius $R_e$, the radius that encloses half the total light of the \sersic\ light profile (\citealt{Sersic.1968}; see also \citealt{Trujillo.2001,Ravindranath.2004,Ravindranath.2006}), is independent of the aperture adopted, although it is subject to the parametric assumption of the \sersic\ profile.

\textit{In this work we adopt the effective half-light radius $R_e$ of a \sersic\ profile along the semi-major axis (which encloses half the total flux) as our definition of galaxy size.} The \sersic\ profile can be expressed in the following equation:
\begin{equation} 
I(R) = I_e \exp \left\{  -b_n\left[\left(\frac{R}{R_e} \right)^{1/n} -1 \right] \right\} 
\end{equation}
where $R_e$ is the effective half-light radius within which half the total flux is enclosed, $I_e$ is the surface brightness at the effective half-light radius, $n$ is the \sersic\ index which characterizes the overall profile shape, and $b_n$ is the normalization constant that ensures that $R_e$ encloses half the total flux (see also \citealt{Graham.2005,Ryan.2010}). Several familiar light profiles such as Gaussian profile ($n=0.5$), exponential profile ($n=1$), and the classical de Vaucouleurs profile ($n=4$; \citealt{deVaucouleurs.1948}) are special cases of the \sersic\ profile. Using a parametric form to estimate galaxy size has the advantage of being independent of the choice of aperture, taking into account the effect of the PSF, and allowing one to weigh individual pixels according to their $S/N$, resulting in more optimal photometry. However it is subject to the assumption of profile shape, which may not always be a good description of real galaxies. We will show through simulations that using the \sersic\ profile-defined galaxy size is less biased than the SExtractor-measured half-light radius in Section \ref{sec:modeling}, at least for galaxies with smooth profiles. We choose GALFIT v3 (\citealt{Peng.2010}) for size and magnitude measurements. GALFIT is a 2D galaxy light-profile fitting software that derives the best-fit parameters of the input functional form that matches the observed galaxy light distribution. GALFIT has been demonstrated to give reasonably unbiased size and magnitude measurements down to the signal-to-noise ratio range typical for high-$z$ galaxies (\citealt{Haeussler.2007,Ravindranath.2006}). 

The summary of our procedure is as follows:
\begin{itemize}
\item We only use one \sersic\ component to fit each galaxy. Although several observational studies (e.g. \citealt{Ravindranath.2006,Law.D.2007,Guo.2012}), as well as studies of galaxy evolution models (e.g. \citealt{Dekel.2009a,Agertz.2009}), have shown that a fraction of high-$z$ star-forming galaxies are clumpy,  we do not use multiple \sersic\ components to obtain better fits. Additional \sersic\ components make it much harder to model the systematic effects, and they do not help us gain more insight on the true size of the galaxies.

\item We use GALFIT to measure the size and magnitude in the passband
that is closest to rest-frame 1500\AA\ in the UV. This means we use the
observed \acsi-band for \acsb-dropouts and the observed \acsz-band for 
\acsv-dropouts.

\item We adopt SExtractor attributes of each dropout galaxy as initial
guesses in GALFIT. For example, we take SExtractor \texttt{MAG\_AUTO} as
the initial guess of total magnitude and half-light radius as the
initial guess of $R_e$. GALFIT is robust enough that it is not very
sensitive to the initial guesses as long as they are reasonably close to
the true values (\citealt{Haeussler.2007}).

\item We make cutouts around every dropout galaxy and run GALFIT on the
cutout images. We expand the cutout image to include close neighbors as
well as bright neighbors that could perturb the fitting result of the
main galaxy. Neighbors are then modeled as separate \sersic\ components.
GALFIT is able to fit multiple \sersic\ components simultaneously, so 
including the close neighbors helps GALFIT obtain more reasonable fits for the
objects of interest. 
 
\item We do not use the dropouts for which GALFIT fails to converge or returns magnitudes fainter than 26.5 mag in GOODS and 28.5 mag in the HUDF. Therefore we do not include the 128 \acsb-dropouts and the 44 \acsv-dropouts that have GALFIT-returned magnitudes fainter than the above limits. We also only include galaxies with acceptable GALFIT results. The fitting quality is determined by both the ratio between $R_e$-error and $R_e$ ($[R_e]_{err}/R_e$) and the reduced chi-square ($\chi^2_\nu$) reported by GALFIT. Therefore we do not include dropouts that have either $[R_e]_{err}/R_e \geq 0.6$ or \acsi-band $\chi^2_\nu > 0.4$, and \acsz-band $\chi^2_\nu > 0.5$ in the GOODS images, and either $[R_e]_{err}/R_e \geq 0.6$ or $\chi^2_\nu > 5.0$ (in both \acsi\ and \acsz) in the HUDF images in the fitting process. (The difference in the values of $\chi^2_\nu$ adopted for GOODS and HUDF images likely originates from the different levels of correlated noise between the two datasets. The median $\chi^2_\nu$ for GOODS and HUDF are 0.33 and 1.22 as derived from simulations, respectively, and the $\chi^2_\nu$ limits correspond roughly to the $90\%$ limit in both fields.) Doing so rejects another 203 \acsb-dropouts and 141 \acsv-dropouts, leaving us with 732 \acsb-dropouts and 280 \acsv-dropouts in the final fitting. 

\end{itemize}

Let us look more carefully at the ÒGALFIT completeness,Ó or the fraction of sources for which GALFIT is able to return good fits. Clearly, at some level, the fact that real LBGs do not always have smooth \sersic-profile morphologies must be contributing to the failure rate. However, as described below, we find that the failure rate in our smooth-profile simulations is very similar to the failure rate for the real galaxies. So we do not think that the morphology difference is the primary cause of GALFIT failures. Instead, we attribute the failure rate to a variety of other factors, such as the influence of neighbors, or problems with centering or background subtraction. Examination of individual failures in the simulations did not reveal obvious trends that we could easily address by tweaking the GALFIT parameters or the fitting procedure.  

\begin{figure}[t]
   \epsscale{1.2}
   \plotone{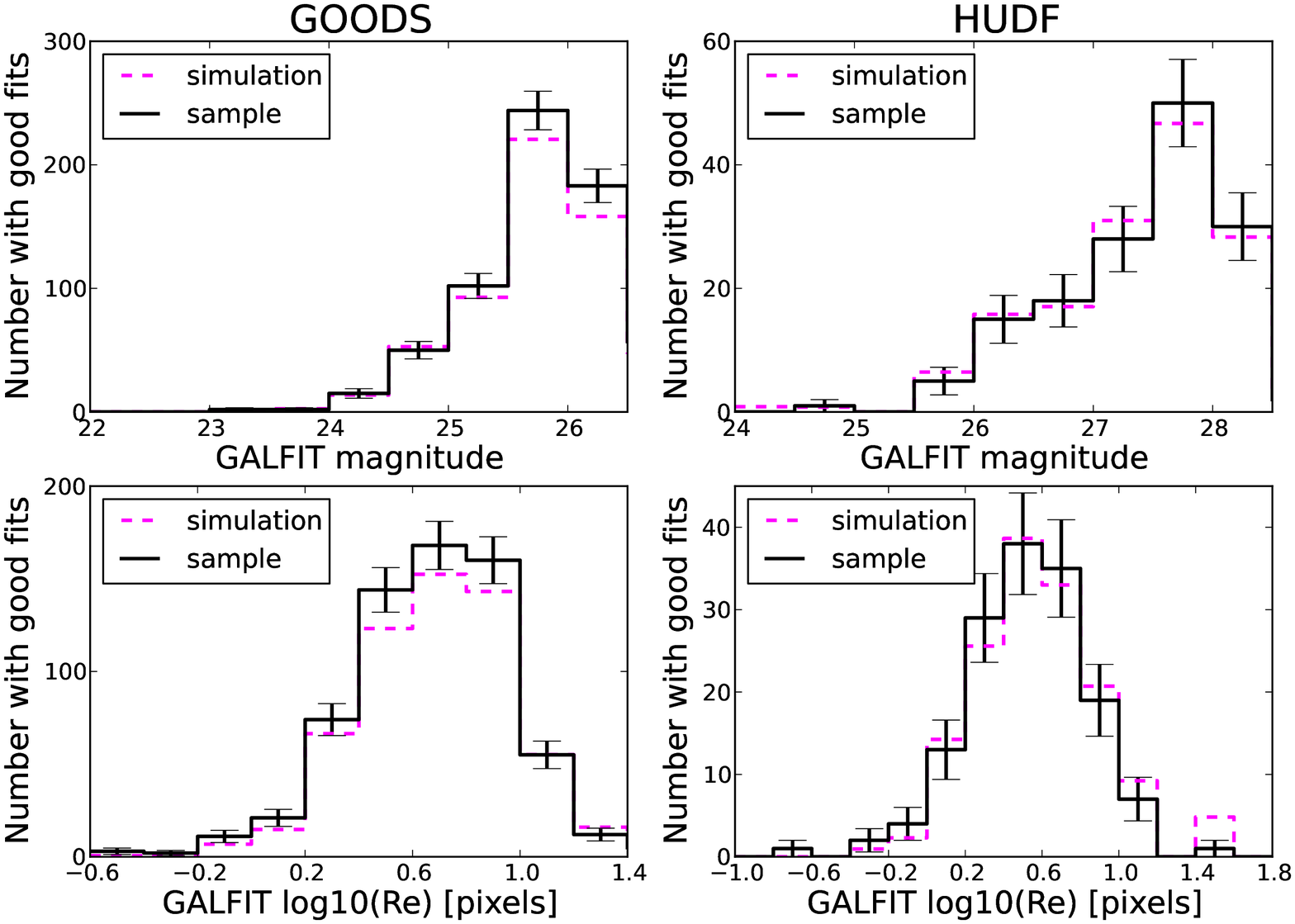}
   \figcaption{The number of sources for which GALFIT is able to return good fits in bins of GALFIT-measured magnitude (top row) and size (bottom row) for \acsb-dropouts. The numbers for GOODS-depth data are shown in the left column; the numbers of HUDF-depth data are shown in the right column. The solid lines show the numbers of \acsb-dropouts; the dashed lines show the numbers expected from simulations. The expected numbers of galaxies with good fits (dashed lines) are calculated by first multiplying the numbers of \acsb-dropouts (whether they have good fits or not) in each magnitude-size bin by the GALFIT completeness of that bin, and then they are marginalize over size (or magnitude) to obtain the expected number of good fits in each magnitude (or size) bin. In each panel we also show the Poisson errors of \acsb-dropouts. In most magnitude or size bins, the expected numbers from simulations are consistent with the numbers of \acsb-dropouts within Poisson errors; however, for the GOODS-depth data the observed numbers are systematically higher than the expected numbers from simulations for magnitudes fainter than 25.5 mag and $\log_{10}(R_e)$ between 0.4 and 1. In these bins the differences are $\sim 10\%$ of the numbers of \acsb-dropouts with good fits. On the other hand, no clear systematic offsets are observed for the HUDF-depth data. The plots for \acsv-dropouts (not shown here) are qualitatively similar and there are no clear systematic offsets for \acsv-dropouts between the expected and the observed numbers with good fits.\label{fig:galfit_qf}}
\end{figure}

In order to understand how well we understand the GALFIT completeness of our sample, and the systematic effects it might have on our number counts, we examine the failure rates for the simulations described in Section \ref{subsec:galfitsim}. In Figure \ref{fig:galfit_qf} the number of \acsb-dropouts with good fits in bins of GALFIT-measured magnitude and size as well as the expected number of good fits determined from our simulations. The left-hand column shows the numbers for the GOODS-depth data (top left is for magnitude bins and bottom left is for size bins); the right-hand column shows the numbers for the HUDF-depth data (top right is for magnitude bins and bottom right is for size bins). The solid lines show the numbers of \acsb-dropouts in each bin; the dashed lines show the expected numbers based on the GALFIT completenesses from simulations. We calculate the expected numbers of galaxies with good fits by multiplying the number of galaxies in each (GALFIT-measured) magnitude-size bin by the GALFIT completeness (derived from simulations) in that bin. Then the total number in a magnitude bin is the sum of the numbers in all bins with the same magnitude but different sizes (the numbers in size bins is calculated in a similar way). The error bars correspond to Poisson errors ($\sqrt{N}$). One can see that in most bins the numbers of observed \acsb-dropouts with good fits are consistent with the expected numbers of good fits from simulations within the errors. However, in the GOODS-depth data there is a slight systematic offset that the observed numbers of \acsb-dropouts with good fits are higher than those expected from simulations, for bins with magnitude fainter than 25.5 mag and $\log_{10}(R_e)$ between 0.4 and 1 (where $R_e$ is in pixels); in those bins, the differences in the numbers is $\sim 10\%$ of the numbers of \acsb-dropouts with good fits. The same offsets are not observed for the HUDF-depth data. It is unclear how much systematic error the offsets introduce into our best-fit size-luminosity distribution, but we do not think that it drastically changes the main conclusions of this paper. 

It appears that fitting \acsb-dropouts and \acsv-dropouts is really pushing the limits of GALFIT. Because we model the completeness due to GALFIT measurements also in magnitude and size bins, these effects will be included in our transformations and will also be reflected in the final results.

\begin{figure}[t]
   \includegraphics[width=\columnwidth]{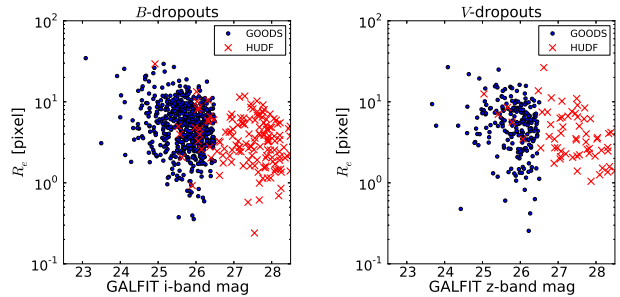}
   \figcaption{GALFIT-measured effective half-light radii and magnitudes of the \acsb- and \acsv-dropout galaxies. Only the sources whose GALFIT results satisfy the criteria described in Section \ref{sec:measure} are shown. Dark points are dropouts from the GOODS dataset; gray crosses are dropouts from the HUDF dataset. The pixel scale is 0.03". The median of the magnitude errors reported by GALFIT of the \acsb-dropouts selected from GOODS and HUDF are 0.31 and 0.20 mag, respectively, while the median errors in $R_e$ (returned by GALFIT) for \acsb-dropouts selected from GOODS and HUDF are 0.5 and 0.3 pixels, respectively. Similarly, the median magnitude errors for \acsv-dropouts selected from GOODS and HUDF are 0.55 and 0.42 mag, respectively, while the median errors in $R_e$ for \acsv-dropouts selected from GOODS and HUDF are 1.4 and 0.8 pixels, respectively. However, the errors are a strong function of surface brightness, so the above errors are not representative of all galaxies in our sample. We will derive the measurement errors in each magnitude-and-size bins with simulations in Section \ref{subsec:galfitsim}. Please see the online version for the color figure.\label{fig:gfmeasure}}
\end{figure}

The GALFIT-measured effective half-light radii and magnitudes for our $B$-dropout and $V$-dropout samples, excluding poor fits, are shown in Figure \ref{fig:gfmeasure}. Some objects have very small best-fit $R_e$, in some cases $R_e < 1$ pixel. Most of them lie very close to the selection limit in the given dataset (\acsz$=26.5$ mag in GOODS and \acsz$=28.5$ mag in HUDF) where rejection using \texttt{CLASS\_STAR} is not reliable, and from visual inspection all of them are unresolved. We exclude known stars and AGNs in the bright end to the best of our knowledge, and some bright sources are genuine compact, high-redshift sources. One example is GOODS J033218.92-275302.7 in the \acsv-dropout sample that has $m_z = 24.43$ mag and $R_e=0.0144"$ (or 0.48 pixel). This object has measured spectroscopic redshift $z_{spec}=5.563$ (\citealt{Vanzella.2009}) and is one of the brightest and the most compact galaxies in our sample.

\begin{figure}[t]
	\epsscale{1.1}
   \plotone{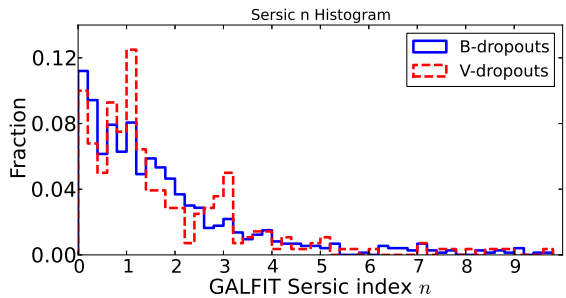}
   \figcaption{The distribution of GALFIT-measured \sersic\ index $n$. The distribution of \acsb-dropouts is shown in  dark solid histogram, while the distribution of \acsv-dropouts is shown in gray dashed histogram. The fraction of late-type galaxies (defined as those having $n \leq 2.5$) is 78\% among \acsb-dropouts and is 72\% among \acsv-dropouts. This motivates us to use 70\% exponential-profile and 30\% de Vaucouleurs-profile objects in the simulations described in Section \ref{subsec:galfitsim}. \label{fig:nhist}} 
\end{figure}

The distribution of GALFIT-measured \sersic\ index for our samples is shown in Figure \ref{fig:nhist}. It is customary to classify the morphology of galaxies using the best-fit \sersic\ index $n$: galaxies with low $n$ (usually $n \leq 2.5$) are classified as ``late-type'', or ``disk-like'', while galaxies with high $n$ (usually $n > 2.5$) are classified as ``early-type'' or ``bulge-like''. Using this definition, among our \acsb-dropout sample 78\% is classified as late-type, while among our \acsv-dropout sample 72\% is classified as late-type. This is consistent with earlier studies (e.g. \citealt{Ravindranath.2006}) that the majority of the LBGs have \sersic\ indices of late-type (or disk-like) galaxies. Since we do not constrain the \sersic\ index, a small fraction of objects have very large \sersic\ indices. The fraction of objects with extremely large $n$ ($n > 8$) is small --- only about 3\% --- and should not affect our results dramatically.

\begin{figure}[t]
   \includegraphics[width=\columnwidth]{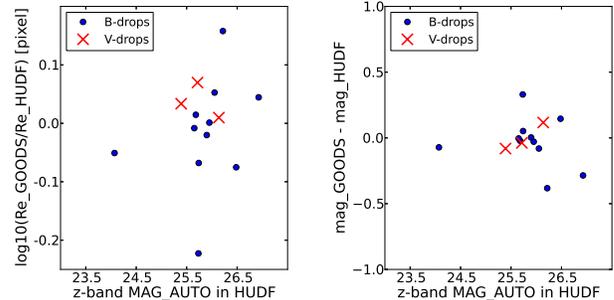}
   \figcaption{Comparison between the GALFIT-measured size (left panel) and magnitude (right panel) in GOODS and HUDF for dropouts detected in both images. The dots are the \acsb-dropouts and the crosses are the \acsv-dropouts. The sizes measured in both images generally agree within 0.1 dex. \label{fig:overlap}}
\end{figure}

When a given source detected in GOODS is also detected in HUDF, we use the measurements in the HUDF. We compare the measurements of these sources from the GOODS images and the HUDF images in Figure \ref{fig:overlap}. 14 \acsb-dropouts and 3 \acsv-dropouts are detected in both images, among them GALFIT failed to return any result for 1 \acsb-dropout and did not return a good fit for 5 \acsb-dropouts. For the dropouts that return good fits, there is good agreement between the two measurements. The GALFIT-measured sizes all lie within $\pm$0.1 dex from each other except for one source.

\section{Modeling and Fitting Procedures}\label{sec:modeling}

In this section we describe the procedure to build the model size-luminosity distribution in detail and much of this section is rather technical. Here we highlight that the new feature of our work is to account for systematic effects of dropout selection and size measurement in magnitude \textit{and} size bins. The steps involve:
\begin{enumerate}
\item\label{step1} constructing a theoretical 2D model that combines the luminosity function and size distribution; 
\item\label{step2} transforming the distribution from step \ref{step1} using the dropout-selection completenesses and redshift distributions (derived in each intrinsic absolute magnitude and effective half-light radius bin) to account for sample selection effects;
\item\label{step3} transforming the distribution from step \ref{step2} using the GALFIT-measurement transfer function kernels (derived in apparent magnitude and effective radius bins) to account for measurement bias and uncertainties; 
\item\label{step4} using the transformed model from step \ref{step3} to compare with the observed size-luminosity distribution and derive the best-fit parameters. 
\end{enumerate}

\subsection{Bivariate Size-Luminosity Distribution}\label{subsec:bivmodel}
In this study we combine the UV luminosity function and size distribution to form the bivariate size-luminosity distribution in rest-frame 1500\AA. We adopt the  Schechter function (\citealt{Schechter.1976}) as our model luminosity function, as is customary in most high-$z$ UV LF studies and was shown to be a decent parametrization (e.g.
\citealt{Steidel.1999,Bouwens.2007,Reddy.2009}):
\begin{equation} \label{eq:schechter} 
   \phi(L)dL = \phi^*
   \left(\frac{L}{L^*}\right)^{\alpha}\exp\left(-\frac{L}{L^*}\right)\frac{dL}{L^*}
\end{equation}
where $\phi(L)dL$ is the volume number density of galaxies within the luminosity range $(L,\,L+dL)$, $L^*$ is the characteristic luminosity, and $\alpha$ is the power-law slope at the faint end, and $\phi^*$ is the normalization. 

For the size distribution of LBGs at a given luminosity, we adopt the log-normal distribution:
\begin{equation} \label{eq:lognormal}
   p(R_e)dR_e=\frac{1}{\sigma_{\ln R_e}\sqrt{2\pi}}
   \exp\left[-\frac{\ln^2(R_e/\bar{R_e})}{2\sigma_{\ln R_e}^2}\right]\frac{dR_e}{R_e}
\end{equation}
where $p(R_e)dR_e$ is the probability density that a galaxy at luminosity $L$ has effective half-light radius between $(R_e,\,R_e+dR_e)$, and $\bar{R_e}$, $\sigma_{\ln \mathrm{R_e}}$ are the peak and width of the distribution, respectively. The log-normal form of size distribution is motivated by the disk-formation theory of \citet{Fall.Efstathiou.1980}, in which gas approximately conserves its specific angular momentum when it forms in the centers of dark matter halos, and disk size is proportional to the spin parameter $\lambda$ and the halo virial radius $R_{\mathrm{vir}}$: $R_d\propto \lambda\,R_{\mathrm{vir}}$. In this picture, the size distribution of the disks should reflect the underlying $\lambda$ distribution of the halos, which is found to be well approximated by a log-normal distribution independent of redshift or halo mass (\citealt{Barnes.1987,Warren.1992,Bullock.2001}). 

To connect equations (\ref{eq:schechter}) and (\ref{eq:lognormal}), we follow the arguments of \citet{deJong.2000} (see also \citealt{Choloniewski.1985}) that adopts a power-law relation to connect the luminosity $L$ and the peak of the size distribution $\bar{R}$:
\begin{equation} \label{eq:powerlaw}
   \bar{R_e}(L) = R_0\left(\frac{L}{L_0}\right)^\beta.
\end{equation}

The peak of the size distribution $\bar{R}$ at a nominal luminosity $L_0$ is defined as $R_0 \equiv \bar{R_e}(L_0)$. We choose $L_0$ to correspond to $M_0=-21$, or $L_{\nu} = 9.12\times 10^{-12}$ erg\ s$^{-1}$\ Hz$^{-1}$ at 1500 \AA. Using Eq. \ref{eq:powerlaw} to connect Eq. \ref{eq:schechter} and Eq. \ref{eq:lognormal}, and expressing the distribution in terms of absolute magnitude $M$, we come to the full bivariate expression:

\begin{equation} \label{eq:bivariate}
  \begin{aligned}
    \Psi(&M,\log_{10} R_e)\,dM\,d\log_{10} R_e =0.4\ln(10)\times\\
       &10^{-0.4(\alpha+1)(M-M^*)}\exp\left[-10^{-0.4(M-M^*)}\right]\times \\
    &\frac{\ln(10)}{\sigma_{\ln R_e}\sqrt{2\pi}} \exp\left\{ \frac{-\left[
    \log_{10}(R_e/R_0)+0.4\beta(M-M_0) \right]^2}{2\sigma_{\ln R_e}^2/\ln^2(10)}
    \right\}\times\\
    &dM\,d\log_{10} R_e.
  \end{aligned}
\end{equation}

\begin{figure}[t]
   \epsscale{1.2}
   \plotone{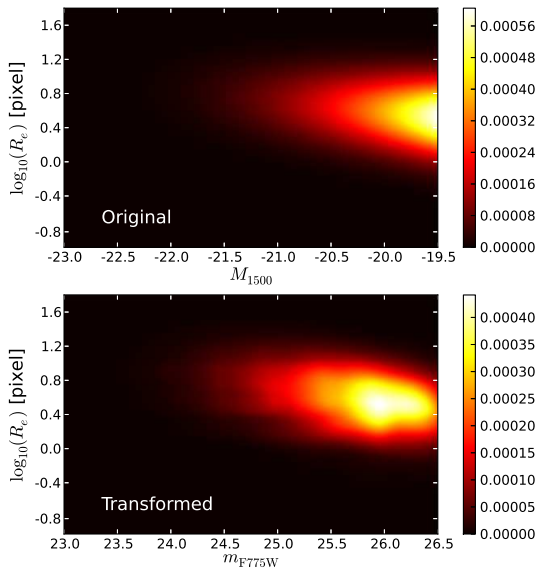}
   \figcaption{This is a visual representation of the size-luminosity model before and after the transformation. The model shown here is parameterized by equation \ref{eq:bivariate} with parameters $\vec{\mathbf{P}}\equiv\params = (-1.7, -21.0, 0.21", 0.7, 0.3)$ down to $m=26.5$ mag --- the adopted magnitude limit of the sample from GOODS. The locations with brighter shades denote higher number densities. We choose the values $\vec{\mathbf{P}}_0$ because these are the initial guesses of the parameters in the fitting process and they are close to the best-fit parameters found later. The top panel shows the model distribution without the corrections of selection effects and measurement biases and uncertainties. The bottom panel shows the distribution after accounting for the aforementioned systematic effects derived from the GOODS-depth \acsi-band image. The mathematical transformations applied will be described in more detail in Section \ref{subsec:dropoutsim} and \ref{subsec:galfitsim}.
\label{fig:twomodels}}
\end{figure}

In Eq. \ref{eq:bivariate}, the bivariate distribution is characterized by a set of five free parameters $\vec{\mathbf{P}} \equiv \params$ (again we determine $\phi^*$ after we determine the shape of the distribution). We show in Figure \ref{fig:twomodels} the distribution in the $(M, \log_{10} R_e)$ plane corresponding to the fiducial parameters $\vec{\mathbf{P}}_0 = (-1.7, -21.0, 0.21", 0.7, 0.3)$ down to apparent magnitude of 26.5 mag. The regions with brighter shades in Figure \ref{fig:twomodels} denote higher number densities. We choose the values $\vec{\mathbf{P}}_0$ because these are the initial guesses of the parameters in the fitting process. If the intrinsic distribution of the galaxy population is indeed described by Eq. \ref{eq:bivariate} with $\vec{\mathbf{P}}_0$, one would expect the observed points to follow the \textit{top} panel of Figure \ref{fig:twomodels} \textit{if} the sample is complete and the measurement error is zero. However there are always sample incompleteness and there are always measurement errors. In the following sections we detail the procedure we use to incorporate sample incompletenesses and measurement biases and errors into our model. The transformed distribution is shown in the bottom panel of Figure \ref{fig:twomodels}, which is the expected distribution if the intrinsic distribution is characterized by $\vec{\mathbf{P}_0}$. We compare the transformed distribution with the observed distribution to derive the best-fit parameters.

\subsection{LBG Selection Function and Dropout Simulations}\label{subsec:dropoutsim}
\subsubsection{Description of Simulation}\label{subsubsec:dropsim_description}

The first set of simulations (``dropout simulations'') are Monte Carlo simulations in which we insert artificial galaxies in the images and try to select them in our LBG sample. We make usual assumptions on the SED of the input sources in order to mimic the high-$z$ LBG population (\citealt{Yan.2005,Verma.2007,Pentericci.2007,Yabe.2009,Stark.2009}), similar to the ones used in the simulations in \citet{Giavalisco.2004b}. However we still describe the simulations in detail here: the input SED is taken from the model of \citet{BC2003} with constant star-formation history of duration 140 Myr; the rest-frame 1500 \AA\ absolute magnitudes are drawn from a flat distribution between $-25$ and $-15$ mag; the redshift is drawn from a flat distribution between 3 and 6; the dust, parameterized by $E(B-V)$, is drawn from a Gaussian distribution with peak 0.15 and dispersion 0.05. These values correspond to the range of rest-frame UV slope ($\beta$) between $-2.3$ and $-0.8$, consistent with the measured range in the literature (see e.g. \citealt{Bouwens.2009,Finkelstein.2012,Castellano.2012}). We use the \citet{Calzetti.2000} extinction law to attenuate the spectrum; we also use the \citet{Meiksin.2006} recipe to model cosmic neutral-hydrogen attenuations. The neutral-hydrogen Lyman-series optical depths of \citet{Meiksin.2006} are smaller than the earlier estimates from \citet{Madau.1995}, but for our filter set and redshift of interest the differences in color corrections are fairly small. Only at the high-redshift end of the \acsv-dropout sample do the two corrections differ appreciably, with the corrected colors using \citet{Madau.1995} redder by $\sim 0.2$ mag. We expect that adopting \citet{Meiksin.2006} will result in lower selection completeness for the \acsv-dropout sample, but a detailed comparison is outside the scope of this work. With the corrected SED, we calculate the magnitudes in \acsb-, \acsv-, \acsi-, and \acsz-bands as our input values.

We use the IRAF package \texttt{artdata} to generate images of artificial galaxies and insert 40 galaxies at a time into a 4096-by-4096-pixel cutout. The input magnitudes range from 22 to 27 mag for the GOODS fields, and 24 to 29 mag for the HUDF. The input effective half-light radii range from 0.05" to 1.5". Each object can have one of two profile shapes --- the exponential profile and de Vaucouleurs profile --- and we assign each object a position in the image and position angle drawn randomly from a flat distribution. We draw an axial ratio for each source from the model distribution  in \cite{Ferguson.2004} that match the observed distribution of their $z\sim 4$ sample. We use 70\%\ exponential and 30\%\ de Vaucouleurs profile as observed in \cite{Ravindranath.2006} (see also Figure \ref{fig:nhist}). In order to properly populate the parameter space in magnitude, effective half-light radii, and redshift, we iteratively run the simulations until we have collected at least 500,000 sources. We derive one set of dropout-selection function $P(M, R_e, z)$ from the output simulation catalog for each dropout sample in each field; i.e., we have two sets of $P(M, R_e, z)$ for the \acsb-dropout sample (one for the GOODS-depth images and one for the HUDF-depth images) and two sets for the \acsv-dropout sample.

The dropout simulations involve many assumptions on the intrinsic SED (e.g. dust attenuation) and the structural parameters (e.g. profile shape and axial ratio) of galaxies representative of the LBG population. Because we do not know the intrinsic distributions of the properties of the entire population, various assumptions are inevitable. These assumptions can only be verified with better constraints on the physical properties of LBGs in the future.

\subsubsection{Calculating Selection Completeness}\label{subsubsec:calc_dropselcomp}

The dropout simulations quantify the LBG selection function $P(M, R_e, z)$ within the absolute-magnitude bin $(M, M+\Delta M)$, effective-radius bin $(\log_{10} R_e, \log_{10} R_e+\Delta \log_{10} R_e)$, and redshift bin $(z, z+\Delta z)$. By definition in each bin of $(M, R_e, z)$,
\begin{equation}\label{eq:selfunc_def}
   P(M, R_e, z) \equiv \frac{\mathrm{\#\ of\ detected\ and\ selected\
sources}}{\mathrm{total\ \#\ of\ input\ sources}}.
\end{equation}
In other words, $P(M, R_e, z)$ is the probability that an LBG in the given $(M, R_e, z)$ bin is detected in the image \textit{and} selected in our sample. We adopt narrow bin sizes of $(\Delta M, \Delta\log_{10} R_e, \Delta z) = (0.5, 0.2, 0.1)$ so that the result is insensitive to the input distribution in either $M$, $R_e$, or $z$ (as long as the input distribution is not pathological).

\begin{figure}[t]
	\epsscale{1.2}
   \plotone{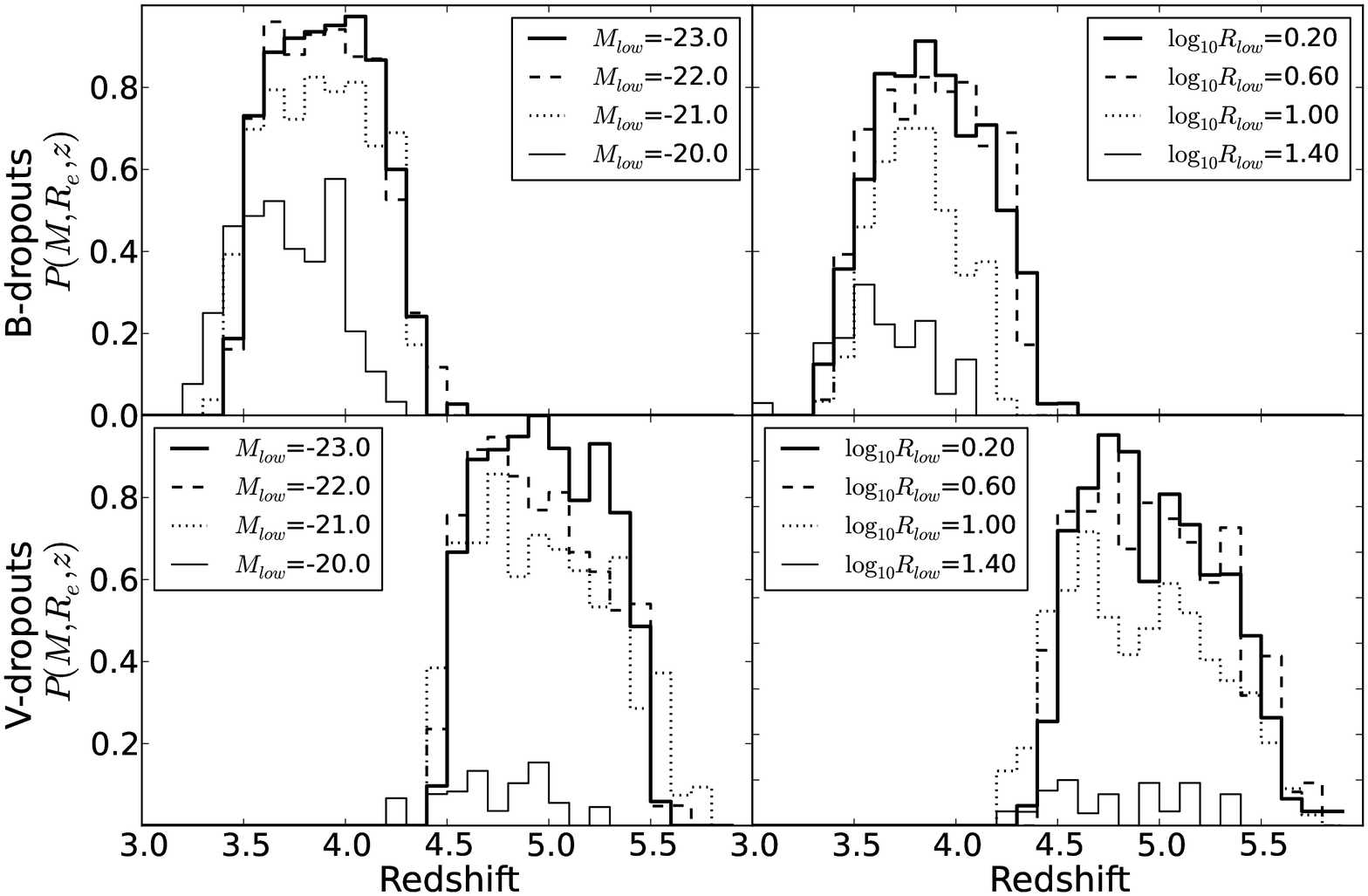}
   \figcaption{The dropout-selection functions $P(M, R_e, z)$ for \acsb-dropouts (top row) and for \acsv-dropouts (bottom row) derived from the GOODS-depth images. In both rows, the function $P(M, R_e, z)$ are shown in different $M$ bins (the left column) and in different $\log_{10} R_e$ bins (the right column). In the left column all bins have $0.6 \leq \log_{10} R_e < 0.8$ ($R_e$ is in pixels of 0.03"), corresponding to roughly $0.12" \leq R_e < 0.19"$. In the right column all bins have the same absolute magnitude $-21\leq M < -20.5$. All absolute magnitudes and effective half-light radii are at rest-frame 1500 \AA. The lower limits in absolute magnitude or effective half-light radius for each curve is indicated in the legend. The units of effective half-light radius is pixels (with scale 0.03"). \label{fig:zdgrid_all}}
   
\end{figure}

Figure \ref{fig:zdgrid_all} illustrates the dependence of the dropout-selection function $P(M, R_e, z)$ on absolute magnitude $M$ and effective half-light radius $R_e$ derived from the simulations in the GOODS-depth images. The top row shows the $P(M, R_e, z)$ for the \acsb-dropout selection; the bottom row shows the same for the \acsv-dropout selection. For both \acsb-dropouts and \acsv-dropouts, the selection function is skewed more towards the low-redshift end in fainter bins as well as in larger-effective-radius bins, showing that the redshift distribution of the dropout sample depends on both absolute magnitude and effective half-light radius. The plots for the HUDF-depth images are qualitatively similar, but they extend to fainter absolute magnitudes.

\subsubsection{Applying $P(M, R_e, z)$}\label{subsubsec:apply_zdgrid}
As the first step towards building the expected observed size-luminosity distribution, we incorporate the dropout-selection incompleteness into equation \ref{eq:bivariate} using $P(M, R_e, z)$. We calculate the expected distribution $\theta$ with the dropout-selection completenesses as
\begin{align}
   \theta(m, R_e) &= \int_{-\infty}^{M_{\mathrm{lim}}}\int_{0}^{\infty} \Psi(M, R_e)\,P(M, R_e, z)\,\frac{dV}{dz}\,\delta_M dMdz\label{eq:theta}\\
   \delta_M &\equiv \delta\left[M-\mathrm{DM}(z)-K-m\right]\label{eq:kcorr}
\end{align}
where $dV/dz$ is the volume element contained in the surveyed area and redshift slice $(z, z+dz)$, and $\delta_M$ is the Dirac delta function that facilitates the transformation from 1500 \AA\ absolute magnitude $M$ to apparent magnitude $m$. DM$(z)$ is the distance modulus  to redshift $z$ and $K$ is the $k$-correction term that converts from 1500 \AA\ to the observed bandpass. The integration limit in absolute magnitude $M_{\mathrm{lim}}$ is chosen to include the faintest dropout candidate possible that could be selected in our sample. Although we integrate over all redshifts, $P(M,R_e,z)$ is only non-zero within the redshift range targeted by the selection criteria. By multiplying the volume element $dV/dz$, the distribution $\theta$ predicts the number counts in the $(m, \log_{10} R_e)$ plane.

When implementing the correction, the integration becomes a discrete sum over small bins of $(M, R_e, z)$ because we only have a discrete sampling of $P(M, R_e, z)$. Therefore we define the bivariate model $\Psi$ on a discrete grid with $(dM, d\log_{10} R_e)=(0.02, 0.02)$. Together with $P(M, R,_e z)$ defined in bins of $(\Delta M, \Delta \log_{10} R_e, \Delta z) = (0.5, 0.2, 0.1)$, we calculate the distribution $\theta$ also on a discrete grid with $(dm, d\log_{10} R_e)=(0.02, 0.02)$. The details of such calculations are elaborated on in Appendix \ref{appendix:selfunc}. 

\subsection{GALFIT Measurement Biases, Uncertainties, and Completenesses}\label{subsec:galfitsim}

\subsubsection{Description of GALFIT Simulation}\label{subsubsec:galfitsim_description}
The second set of Monte Carlo simulations (``GALFIT simulations'') quantifies the measurement biases and uncertainties of magnitude and effective half-light radius. In this set of simulations, we again insert artificial galaxies using IRAF package \texttt{artdata} with the same distributions of magnitude and structural parameters ($R_e$, position angle, ellipticity, etc.) as the dropout simulations (Section \ref{subsec:dropoutsim}). We calculate transfer function kernels in apparent-magnitude and effective-radius bins of width $(\Delta m, \Delta \log_{10} R_e)=(0.5, 0.2)$ and assume that the kernel is unchanged within the bin, and we believe that the chosen bin widths are narrow enough so that the derived kernels should be relatively insensitive to the input magnitude and size distribution. During the simulations we run SExtractor to detect the artificial sources and then use GALFIT to measure their magnitudes and effective half-light radii. Because we only care about how GALFIT performs in the passband closest to rest-frame 1500 \AA, we only collect results in the \acsi-band for the \acsb-dropout sample and in the \acsz-band for the \acsv-dropout sample. We run the process iteratively until we collect at least 100,000 objects in each band.

\subsubsection{GALFIT Transfer Function Kernels}\label{subsubsec:gfkernels}
Using the output simulation catalog, we calculate the differences between the input and output magnitudes and effective half-light radii for all sources. We denote the input magnitude and effective half-light radius as $m$ and $R_e$, and the output (measured) values as $m'$ and $R'_e$. The differences between the input and the measured magnitudes and sizes are defined as $\delta m \equiv m'-m$ and $\delta \log_{10} R_e \equiv \log_{10} R'_e - \log_{10} R_e$, respectively. Then in the $i'$-th magnitude bin (in which $m_{i'} < m < m_{i'}+\Delta m$) and the $j'$-th size bin (in which $(\log_{10} R_e)_{j'} < \log_{10} R_e < (\log_{10} R_e)_{j'} + \Delta\log_{10} R_e$), we calculate the transfer function kernel $T_{i'j'}^\mathrm{GF}$ and completeness $C^{\mathrm{GF}}_{i'j'}$ as
\begin{align}
   T_{i'j'}^{\mathrm{GF}} &\equiv \mathcal{P}(\delta m, \delta \log_{10} R_e)\label{eq:GFkernel}\\
   C^{\mathrm{GF}}_{i'j'} &= \int\,\int\,\mathcal{P}(\delta m, \delta \log_{10} R_e)\,d\delta m\,d\delta \log_{10} R_e  
\end{align}
where $\mathcal{P}(\delta m, \delta \log_{10} R_e)$ is the probability density function (PDF) in the plane of ($\delta m$, $\delta\log_{10} R_e$). The GALFIT-measurement completeness $C^\mathrm{GF}_{i'j'}$ accounts for the fact that GALFIT does not always give satisfactory fits. (Here $C^\mathrm{GF}_{i'j'}$ does not include the SExtractor-detection incompleteness.) We use the same criteria as described in Section \ref{sec:measure} to calculate the fraction of sources poorly fit by GALFIT, namely those with GALFIT-returned errors $[R_e]_{err}/R_e \geq 0.6$ and $\chi^2_{\nu} > 0.4$ in the GOODS-depth images (and $\chi^2_{\nu} > 0.5$ in \acsz-band), and $[R_e]_{err}/R_e \geq 0.6$ and $\chi^2_{\nu} > 5.0$ in the HUDF-depth images in both \acsi- and \acsz-band. 

\begin{figure}[t]
  \includegraphics[width=\columnwidth]{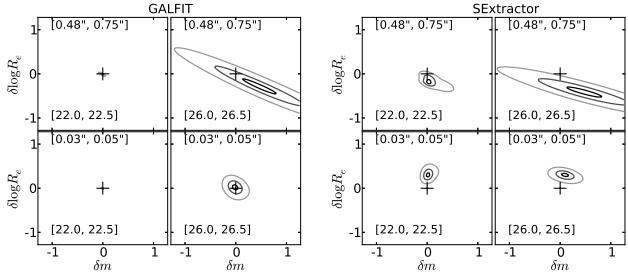}
  \figcaption{The GALFIT (the left half) and SExtractor (the right half) measurement biases and uncertainties in different apparent-magnitude and effective-radius bins, represented by the probability distributions (PDF) in $\delta m$ (measured magnitude $-$ input magnitude) and $\delta \log R_e$ (measured $\log R_e$ $-$ input $\log R_e$). The PDFs (equivalent to the kernels described in Section \ref{subsubsec:gfkernels}) are derived from the GOODS-depth \acsi-band images. The input magnitude range of each bin is in the lower left corner, while the input size range is in the upper left corner. The contours represent 90\%, 50\% and 10\% of the peak value of the distribution. The panels on the right are for fainter magnitude bins, and the panels closer to the top are for larger-size bins. The plus sign denotes the point where $(\delta m, \delta \log_{10} R_e)=(0,0)$. \label{fig:gfsekernels}}
\end{figure}

The transfer function kernels $T_{i'j'}^{\mathrm{GF}}$ in different bins are shown in Figure \ref{fig:gfsekernels}. The panels on the lower left have high surface brightnesses, while the panels on the upper right have low surface brightnesses. In general surface brightness has an important effect on the behavior of the GALFIT measurement, namely kernels in low-surface brightness bins are more spread out (larger uncertainties) and center farther away from the origin (larger biases). The errors in the measured magnitudes and effective half-light radii are correlated as expected: an overestimation of flux ($\delta m < 0$) leads to an overestimation of effective half-light radius ($\delta \log_{10} R_e > 0$). GALFIT performs well overall; even in low-surface brightness bins, GALFIT still shows little bias compared with SExtractor.

\subsubsection{Comparison with SExtractor}\label{subsubsec:sekernels}
One can also calculate the transfer function kernels for SExtractor-measured values. The kernels derived for SExtractor are shown in the right half of Figure \ref{fig:gfsekernels}. One can see the biases of SExtractor-measured values: SExtractor tends to overestimate the effective half-light radius in high-surface brightness bins (lower left) and underestimate the effective half-light radius in low-surface brightness bins (upper right). This is the main reason we decide to use GALFIT instead of SExtractor for size measurement.

\begin{figure}[t]
	\epsscale{1.2}
   \plotone{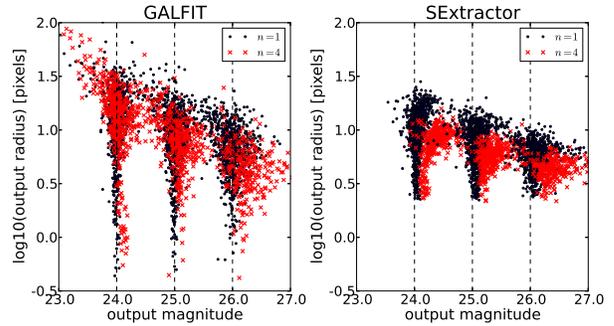}
   \figcaption{The measured magnitudes and sizes for the simulated sources within narrow ranges of input magnitudes around $m=$ 24, 25, and 26 mag derived from GOODS-depth \acsi-band images. We show measurement results for both GALFIT (left panel) and SExtractor (right panel). Results are shown for different input profile shapes: exponential profiles ($n=1$) are shown in dots, while the de Vaucouleurs profile ($n=4$) are shown in crosses. Both software show larger biases for the de Vaucouleurs profile, and the faint, extended wing of this profile is likely the cause. However, SExtractor shows larger biases for both profile shapes.\label{fig:gf_se_bias}}
\end{figure}

We also show the plots of the measured magnitudes and effective half-light radii for very narrow ranges of input magnitudes around $m=$ 24, 25, and 26 mag in Figure \ref{fig:gf_se_bias}. One can also see here that SExtractor in general systematically returns magnitudes that are fainter than the input values. This indicates that SExtractor systematically misses a portion of the light from simulated sources. Although GALFIT measurements show similar scatter, there is no obvious bias in Figure \ref{fig:gf_se_bias}. These comparisons demonstrate the advantages of profile fitting, at least in the case of smooth galaxies. In future efforts, we plan to quantify the biases for realistic --- and possibly clumpy --- galaxies drawn from hydrodynamical simulations.

\subsubsection{Applying $T^{\mathrm{GF}}$}\label{subsubsec:apply_gfkernels}
We apply the GALFIT transfer function kernels $T^{\mathrm{GF}}_{i'j'}$ to $\theta(m, R_e)$ in equation \ref{eq:theta}. This is done by convolving $\theta(m, R_e)$ within the $i'j'$-th bin of $(m, R_e)$ with the transfer function kernel pertinent to that bin, and summing the contribution from all bins together:

\begin{align}
   \psi_{i'j'}(m', \log_{10} R_e) &= \theta_{i'j'}(m, \log_{10} R_e) \ast T^\mathrm{GF}_{i'j'}
   (\delta m,\delta\log_{10} R_e)\label{eq:psi_ij}\\
   \psi(m',\log_{10} R'_e) &= \sum_{i'j'}\,\psi_{i'j'}(m',\log_{10} R'_e)\label{eq:psi}
\end{align}

The distribution $\psi(m', \log_{10} R'_e)$ is the expected distribution in measured magnitude and effective half-light radius corresponding to the set of parameters $\vec{\mathbf{P}}$. Therefore $\psi(m', \log_{10} R')$ accounts for all the systematic effects related to dropout-selection and GALFIT measurement. We define the final distribution on the grid of bin size $(dm', d\log_{10} R'_e) = (0.02, 0.02)$, and the detailed mathematical formulation of Eq. \ref{eq:psi_ij} and \ref{eq:psi} on the grid is included in Appendix \ref{appendix:gfkernel}. After applying the dropout-selection function $P(M, R_e, z)$ (Section \ref{subsubsec:apply_zdgrid}) and GALFIT-transfer-function kernels $T^{\mathrm{GF}}$, we arrive at the expected distribution of observed dropouts. The distribution corresponding to the fiducial set of parameters $\params = (-1.7, -21.0, 0.21", 0.7, 0.3)$ is shown in the bottom panel of Figure \ref{fig:twomodels}.

\subsection{Fitting Procedure}\label{subsec:fitting}
We use the maximum-likelihood estimation (MLE) method to recover the best-fit parameters $\vec{\mathbf{P}}\equiv\params$. Our formulation is equivalent to that proposed by \citet{Sandage.1979} (see also \citealt{Trenti.2008,Su.2011}). In general, given a set of observed data points $(x_1, x_2, ..., x_n)$ and a model $\Gamma(x)$ that describes the distribution of the observed points, the likelihood of the model $\Gamma(x)$ is
\begin{equation}\label{eq:likelihood}
   \mathcal{L} \equiv \Gamma(x_1)\cdot\Gamma(x_2)\cdots\Gamma(x_n)
   = \prod_{i=1}^{n}\Gamma(x_i)
\end{equation}
Here we try to determine the set of parameters $\vec{\mathbf{P}}\equiv\params$ that produces the model $\psi(m',\log_{10} R'_e)$ that will maximize the likelihood function $\mathcal{L}$ for the observed magnitudes and sizes. Here we also need to include the component of the interlopers $I(m', \log_{10} R'_e)$ described in Section \ref{subsec:interloper} (and calculated in Appendix \ref{subsec:interloper_contrib}) that accounts for the number counts of possible low-$z$ interlopers due to photometric scatter. Therefore for each dropout sample we combine the log-likelihood from both fields (GOODS+HUDF) and find the set of parameters that maximizes the combined log-likelihood. In short, for a given set of parameters $\vec{\mathbf{P}}$
we first calculate the overall normalization constant $\phi^*$ through
the equation
\begin{equation}\label{eq:phistar}
\begin{aligned}
      N_{\mathrm{G}} + &N_{\mathrm{H}} = \phi^*\sum_{s,l}\biggl[\psi^{\mathrm{G}}
      (m'_s, \log_{10} (R'_e)_l)+I^{\mathrm{G}}(m'_s,\log_{10} (R'_e)_l)\\
      +&\psi^{\mathrm{H}}(m'_s, \log_{10} (R'_e)_l)+I^{\mathrm{H}}(m'_s,\log_{10} (R'_e)_l)\biggl]dm'd\log_{10}R'_e
\end{aligned}
\end{equation}
where $N_{\mathrm{G}}$ and $N_{\mathrm{H}}$ are the total number of dropouts in GOODS and HUDF, respectively. The volume surveyed in each field has already been included in $\psi(m', \log_{10} R'_e)$ (see Section \ref{subsubsec:apply_zdgrid} and Eq. \ref{eq:theta}). We then calculate the sum of the
log-likelihood of both fields:
\begin{subequations}
   \begin{equation}
   \begin{aligned}
      \log_{10} \mathcal{L}_{\mathrm{G}}(\vec{\mathbf{P}}) &= N_{\mathrm{G}}
      \,\log_{10} \phi^*(\vec{\mathbf{P}})\\
      &+ \sum_{i=1}^{N_{\mathrm{G}}}
      \log_{10} \biggl[\psi^{\mathrm{G}}(\vec{\mathbf{P}};\,m'_{i}, \log_{10} (R_e)_i)\\
      &+I^{\mathrm{G}}(m'_i, \log_{10} (R'_e)_i)\biggl] \label{eq:logL1}
   \end{aligned}
   \end{equation}
   \begin{equation}
   \begin{aligned}
      \log_{10} \mathcal{L}_{\mathrm{H}}(\vec{\mathbf{P}}) &= N_{\mathrm{H}}
      \,\log_{10} \phi^*(\vec{\mathbf{P}})\\
      &+ \sum_{i=1}^{N_{\mathrm{H}}} 
      \log_{10} \biggl[\psi^{\mathrm{H}}(\vec{\mathbf{P}};\,m'_{i}, \log_{10} (R_e)_{i})\\
      &+I^{\mathrm{H}}(m'_i,\log_{10} (R'_e)_i)\biggl] \label{eq:logL2}
   \end{aligned}
   \end{equation}
   \begin{equation}
      \log_{10} \mathcal{L}_{\mathrm{tot}}(\vec{\mathbf{P}}) = \log_{10}\mathcal{L}_{\mathrm{G}}(\vec{\mathbf{P}}) + \log_{10}\mathcal{L}_{\mathrm{H}}(\vec{\mathbf{P}})
   \end{equation}
\end{subequations}

We use the \texttt{scipy} package \texttt{optimize} to find the global maximum of $\log_{10} \mathcal{L}_{\mathrm{tot}}$. Within the package \texttt{optimize}, the function \texttt{fmin\_l\_bfgs\_b} uses the L-BFGS-B algorithm (\citealt{Zhu.1997}) to search for global minimum. The best-fit parameters for each dropout sample are presented in Section \ref{sec:bestfitpar}.

\section{Best-Fit Parameters and Confidence Intervals}\label{sec:bestfitpar}
\begin{deluxetable*}{llllllll}
   \tablecaption{The best-fit parameters of dropout samples. \label{tab:bestfit_params}}
   \tablewidth{\textwidth}
   \tablehead{ \colhead{Sample} & \colhead{$\alpha$} & \colhead{$M^*$} &
      \colhead{$\phi^*$} & \colhead{$R_0$} & 
      \colhead{$R_0$\tablenotemark{a,b}} & \colhead{$\sigma_{\ln R_e}$\tablenotemark{c}} & 
      \colhead{$\beta$}\\
      \colhead{} & \colhead{} & \colhead{} & \colhead{($10^{-3}\,\mathrm{Mpc}^{-1}$)} &
      \colhead{(arcsec)} & \colhead{(kpc)} & \colhead{}}
   \startdata
      \acsb-dropouts  &  $\alpharesb\alpharesberr$  &  $\mstarresb\mstarresberr$  &  
         $\phistarresb\phistarresberr$  &  $\Rzeroresbas\Rzeroresbaserr$  &  
         $\Rzeroresbkpc\Rzeroresbkpcerr$  &  $\sigmaresb\sigmaresberr$  &  
         $\betaresb\betaresberr$\\
      \acsv-dropouts  &  $\alpharesv\alpharesverr$  &  $\mstarresv\mstarresverr$  &  
         $\phistarresv\phistarresverr$  &  $\Rzeroresvas\Rzeroresvaserr$  &  
         $\Rzeroresvkpc\Rzeroresvkpcerr$  &  $\sigmaresv\sigmaresverr$  &  
         $\betaresv\betaresverr$\\
   \enddata
   \tablenotetext{a}{assuming $z=4$ for \acsb-dropouts and $z=5$ for \acsv-dropouts.}
   \tablenotetext{b}{peak of the size distribution at $M_{1500}=-21$ mag}
   \tablenotetext{c}{$\sigma_{\ln R_e}=\sigma_{\log_{10} R_e}\times \ln(10)$, where $\sigma_{\log_{10} R_e}$ is in dex}
\end{deluxetable*}

\begin{figure*}[t]
   \plotone{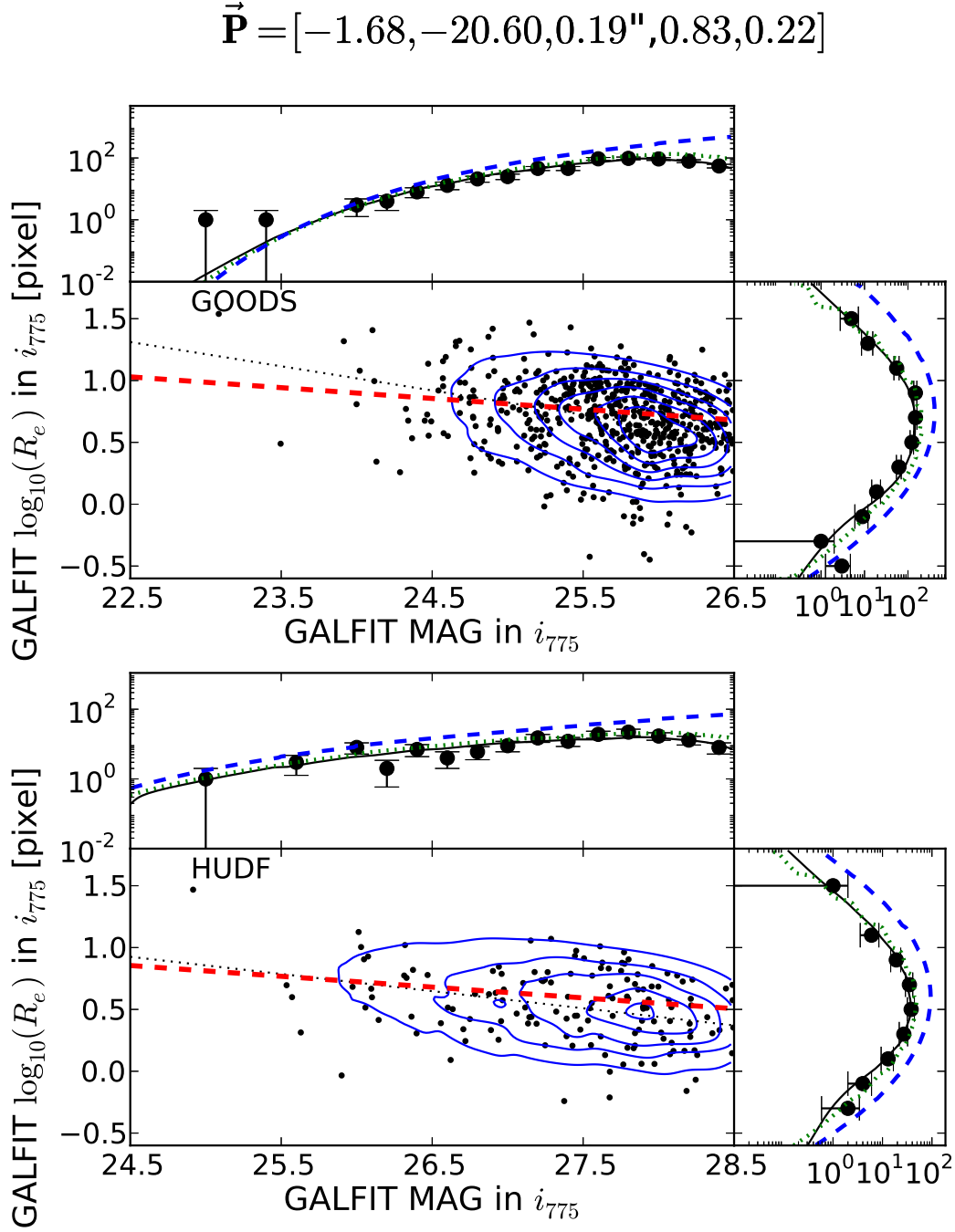}
   \figcaption{The GALFIT-measured magnitudes and effective half-light radii of \acsb-dropouts with best-fit model (with interloper contribution) in the GOODS (upper panel) and in the HUDF (lower panel). The set of best-fit parameters $\vec{\mathbf{P}}$ is shown at the top. The \acsb-dropouts used for fitting are shown as black points in the main panel. The contours correspond to the best-fit model accounted for both dropout-selection and GALFIT-measurement. The thin dotted line is the line that minimizes the sum of the distances to all the data points; the thick dashed line is the best-fit size-luminosity relation. The top inset of each panel shows the marginalized luminosity function from the best-fit model. The dashed line is the Schechter function corresponding to a boxcar selection function within $3.5\leq z \leq 4.5$; the dotted line is the function accounted for dropout-selection effects only; the solid line is the function accounted for both the dropout-selection and the GALFIT-measurement effects. The black points with Poisson error bars are the number of dropouts in each magnitude bin. The right inset of each panel shows the marginalized size distribution. The dashed line is the log-normal distribution; the dotted line is the size distribution accounted for dropout selection; the solid line is the distribution accounted for both dropout selection and GALFIT measurements. The black points with Poisson error bars are the observed numbers in each size bin. \label{fig:bdrops_fit}}
\end{figure*}

\begin{figure*}[t]
   \plotone{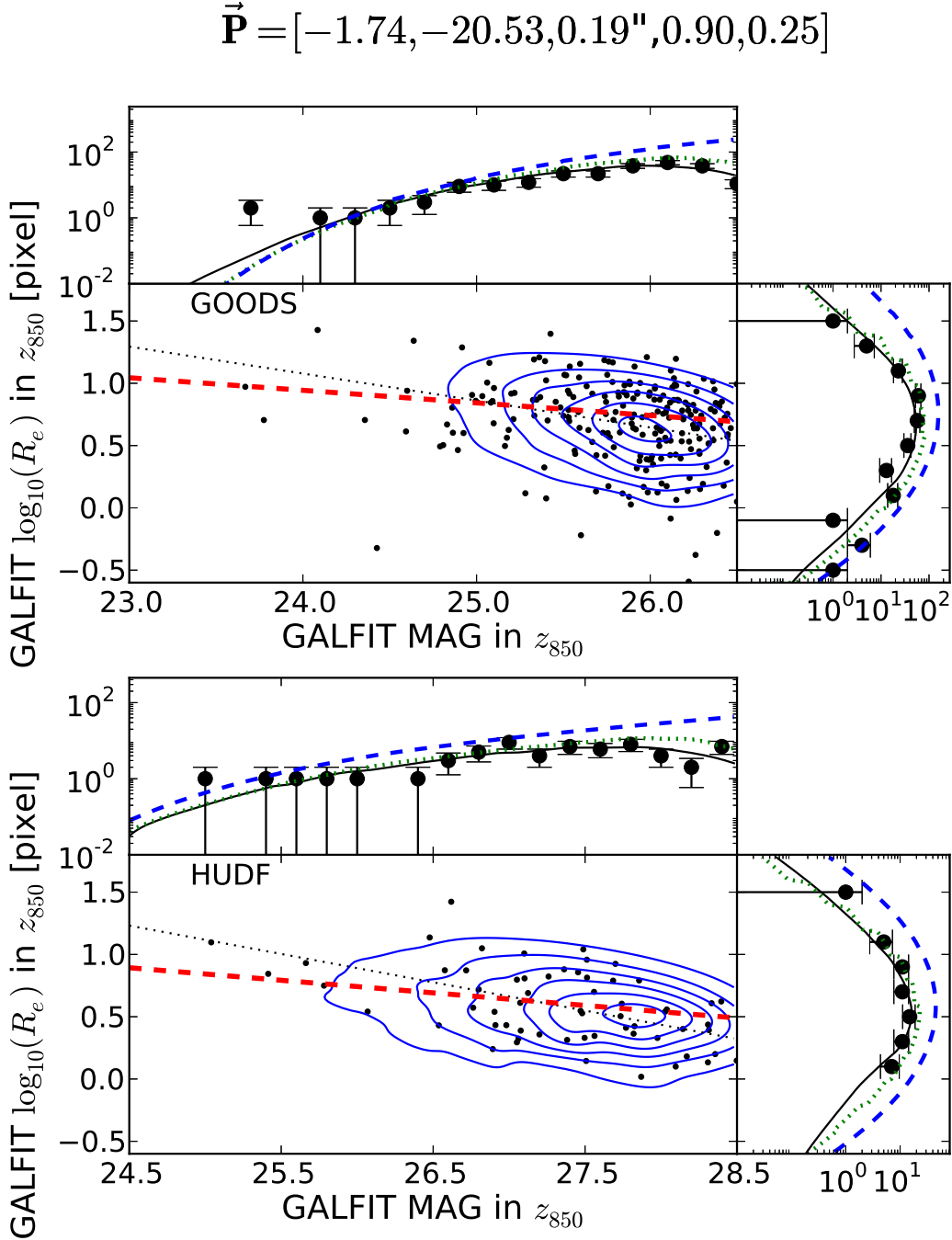}
   \figcaption{The GALFIT-measured magnitudes and effective half-light radii of \acsv-dropouts in GOODS and in the HUDF with the best-fit model. The plot legend is the same as Figure \ref{fig:bdrops_fit}.
\label{fig:vdrops_fit}}
\end{figure*}

The best-fit values of $\fullparams$ for both \acsb-dropouts and \acsv-dropouts are listed in Table \ref{tab:bestfit_params}. We plot the model distributions corresponding to the best-fit values along with the GALFIT-measured magnitudes and effective half-light radii in Figures \ref{fig:bdrops_fit} and \ref{fig:vdrops_fit}. Also plotted are the projections of the best-fit distribution onto the magnitude axis (top panel) and the effective-radius axis (right panel). Though we derive the best-fit parameters using the GOODS and HUDF fields jointly, we plot them separately for clarity.

Although the maximum-likelihood estimator is a powerful tool to obtain the best parameters, it does not tell us how well the best-fit model matches the data. Therefore we adopt a Monte Carlo approach to assess the goodness of fit. We use the best-fit model to randomly draw a simulated data set that contains exactly the same number of points as the observed data. We then calculate the likelihood of the simulated data set and compare it with the likelihood of the real data. We repeat the process 5000 times and calculate the fraction of simulated data sets that have lower likelihood than the real data. For the \acsb-dropout sample $89\%$ of the simulated data sets have lower likelihood than the real data; for the \acsv-dropout sample $42\%$ have lower likelihood than the real data. We conclude that the best-fit models describe the observed distribution reasonably well.

We show two size-luminosity relations in Figures \ref{fig:bdrops_fit} and \ref{fig:vdrops_fit}. The black dotted line minimizes the distance to all the observed points, so it traces the observed distribution. The red dashed line corresponds to the best-fit size-luminosity relation with transfer functions. The slope of the black dotted line is steeper than that of the red dashed line because the latter tries to account for the surface-brightness-related selection effects.

We plot the projected size distribution for both fields in Figure \ref{fig:bdrops_fit} and \ref{fig:vdrops_fit} in the right panels. The dashed curve is the size distribution predicted by the best-fit model parameters; the dotted curve is the distribution after accounting for the dropout-selection incompletenesses only, and the solid curve is the distribution after including GALFIT completeness, bias and scatter. The peak of the predicted intrinsic distribution (the blue dashed curve) is above the PSF size, implying that the decline of size distribution at sizes smaller than the observed peak reflects the true peak of the underlying size distribution, and is not due to measurement incompletenesses alone.  

The predicted intrinsic luminosity function (as parameterized as Schechter function) as well as the transformed luminosity functions are also plotted in the top panels in Figure \ref{fig:bdrops_fit}. The shape of the predicted intrinsic Schechter function and the transformed luminosity function trace each other well, and serious incompleteness only sets in at the very faint end of our observed magnitude range.

\begin{figure}[t]
	\epsscale{1.3}
   \plotone{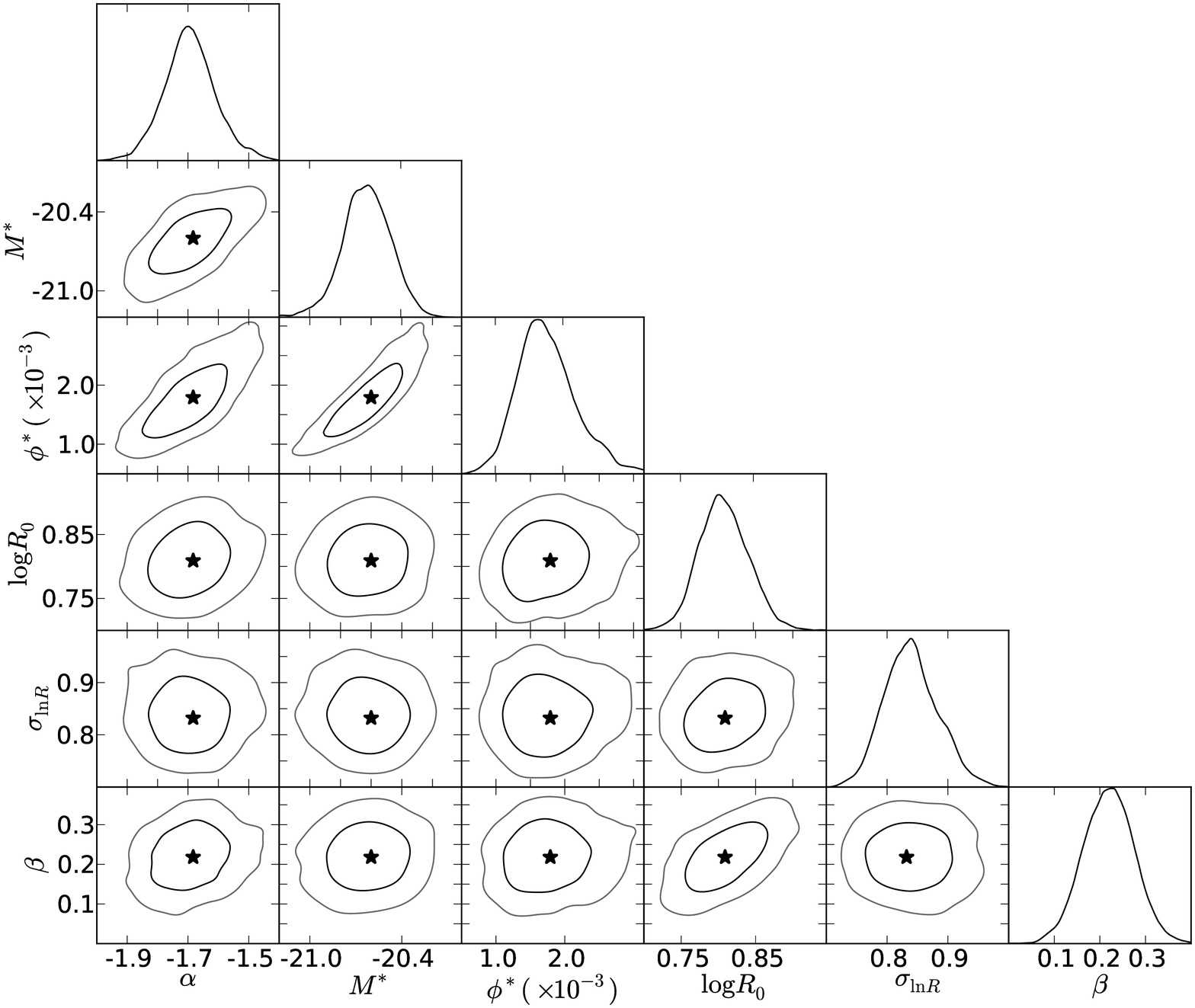}
   \figcaption{The 68\% (dark gray) and 95\% (light gray) confidence intervals of \acsb-dropouts for each pair of parameters. On the top of each column are the normalized probability density function (with arbitrary units on the $y$-axis) of each parameter labeled at the bottom of each column. The unit of $\log R_0$ is in pixels, with pixel scale $0.03"$. The correlations between the LF parameters ($\alpha$, $M^*$, $\phi^*$) and the size-distribution parameters ($\log_{10} R_0$, $\sigma_{\ln R}$, $\beta$) are relatively weak. \label{fig:bdrops_mcmc}}
\end{figure}

\begin{figure}[t]
	\epsscale{1.3}
   \plotone{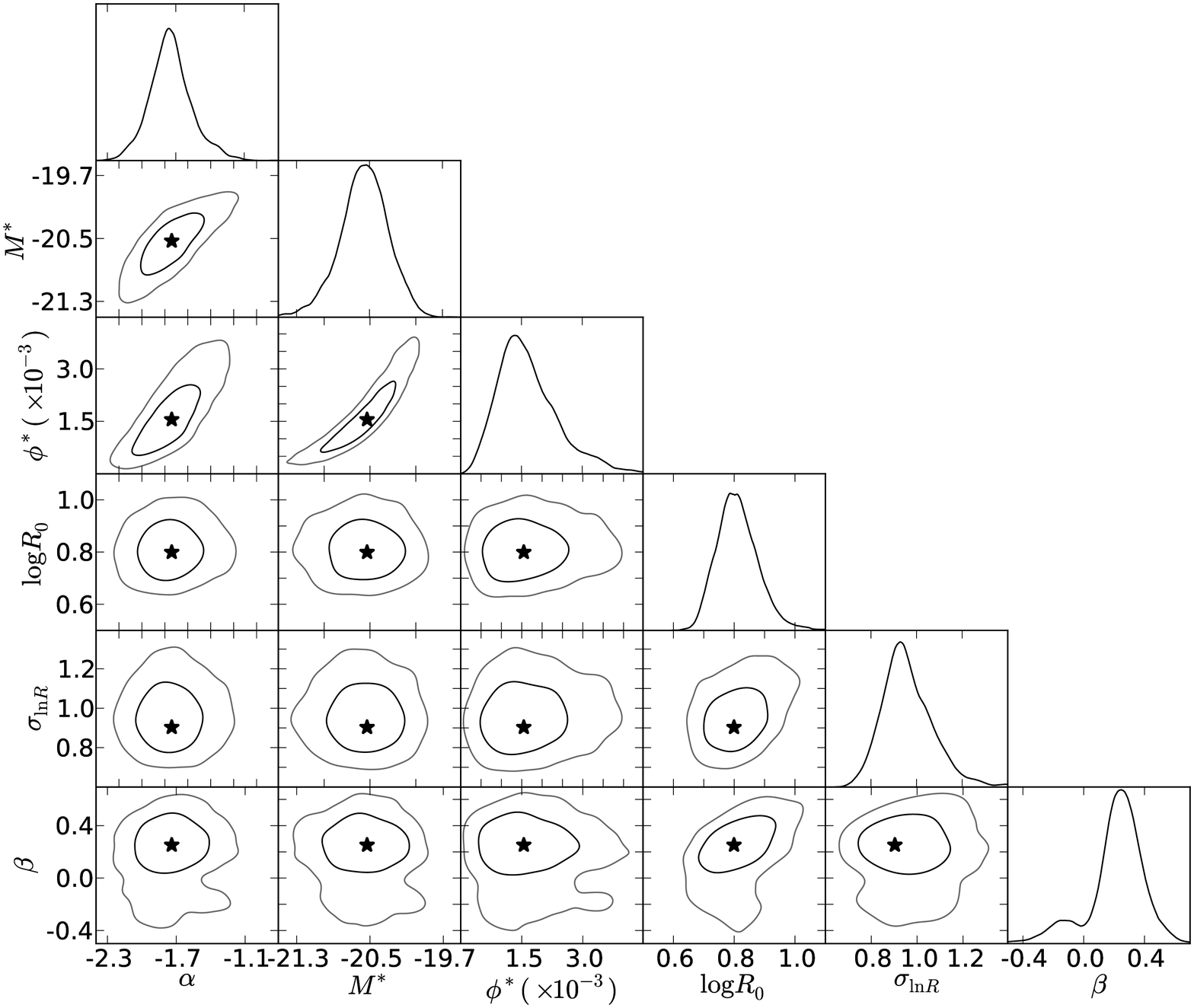}
   \figcaption{The 68\% (dark gray) and 95\% (light gray) confidence intervals of \acsv-dropouts for each pair of parameters. On the top of each column are the normalized probability density function (with arbitrary units on the $y$-axis) of each parameter labeled at the bottom of each column. The unit of $\log R_0$ is in pixels, with pixel scale $0.03"$. Similar to the case of \acsb-dropouts, the correlations between the LF parameters ($\alpha$, $M^*$, $\phi^*$) and the size-distribution parameters ($\log_{10} R_0$, $\sigma_{\ln R}$, $\beta$) are relatively weak. Furthermore, the parameters for the \acsv-dropouts are less well constrained than those for \acsb-dropouts, likely due to a smaller sample size, as the scales on the axes indicate.
   \label{fig:vdrops_mcmc}}
\end{figure}

We use the Markov Chain Monte Carlo (MCMC) sampling to derive the uncertainties for each parameter and to explore the degeneracies among the different parameters. The MCMC method is a powerful way to sample complex, multi-dimensional probability distribution functions numerically. To do this, we use the public python package \texttt{emcee}\footnote{\url{http://danfm.ca/emcee}} (\citealt{Foreman-Mackey.2012}). In Figure \ref{fig:bdrops_mcmc} and \ref{fig:vdrops_mcmc} we show the $68\%$ and $95\%$ confidence contours for all pairs of parameters.  We plot the contours of both samples on the same scale so one can see immediately that the parameters for the \acsv-dropout sample are less well constrained than the \acsb-dropout sample. This is not surprising given the smaller sample size of the \acsv-dropouts. Because the marginal distribution around each best-fit parameter is in general not Gaussian, we quote errors in Table \ref{tab:bestfit_params} as the narrowest intervals around the best-fit values that bracket 68\% of the probability. There is some weak correlation between the luminosity-function parameters ($\alpha$, $M^*$, and $\phi^*$) and the size-distribution parameters ($\log R_0$, $\sigma_{\ln R_e}$, and $\beta$). Especially interesting is the weak correlation between $\alpha$ and $\beta$ (the Pearson's correlation coefficient $r=0.27$): shallower size-luminosity relation (smaller $|\beta|$) leads to steeper faint-end slope $\alpha$ due to larger incompleteness corrections in the faint end. For example, if $\beta=0.1$, the faint-end slope $\alpha$ becomes $\sim -1.9$. A similar correlation was also discussed in \cite{Blanton:2001cp} and \citet{Grazian.2011}. The errors this degeneracy between $\alpha$ and $\beta$ introduces when one derives just luminosity function alone likely requires careful simulations to quantify, but from the contours of Figure \ref{fig:bdrops_mcmc} we estimate that the additional uncertainty in $\alpha$ is likely to be at most $\sim \pm0.05$ for \acsb-dropouts. 

In deriving the corrections for GALFIT measurement (the transfer-function kernels; see Section \ref{subsec:galfitsim}) we make the assumption that the input sources comprise 70\% exponential (late-type) profiles and 30\% de Vaucoulerus (early-type) profiles. This is a reasonable approximation to what we see in Figure \ref{fig:nhist}. However we also tried fitting with kernels derived with 100\% exponential profiles (disk-only kernels) and 100\% de Vaucouleurs profiles (spheroid-only kernels). The best-fit parameters derived using disk-only kernels are indistinguishable from our nominal results: the best-fit parameters for \acsb-dropouts are $\vec{\mathbf{P}}=(-1.68, -20.59, 0.19", 0.82, 0.22)$, while those for \acsv-dropouts are $\vec{\mathbf{P}}=(-1.75, -20.55, 0.19", 0.88, 0.25)$. These are within the 68\% contours of our nominal best-fit parameters. On the other hand, the parameters derived using the spheroid-only kernels are significantly different from our nominal results: the best-fit parameters using the spheroid-only kernels for \acsb-dropouts are $\vec{\mathbf{P}}=(-1.55, -20.33, 0.32", 0.97, 0.11)$, while those for \acsv-dropouts are $\vec{\mathbf{P}}=(-1.43, -20.50, 1.30", 1.50, 0.39)$. While we can not reject the hypothesis that all LBGs have $n=4$ \sersic\ profiles based on the best-fit size-luminosity distributions, it is unlikely given the observed \sersic-index distribution in Figure \ref{fig:nhist}. We conclude that our size-luminosity distributions are reasonable as long as the majority of LBGs have close to exponential profiles.

\section{Discussion}\label{sec:discussion}

\subsection{Comparison with Previous UV LF Determinations}\label{subsec:LFcompare}
\begin{figure}[t]
	\epsscale{1.2}
   \plotone{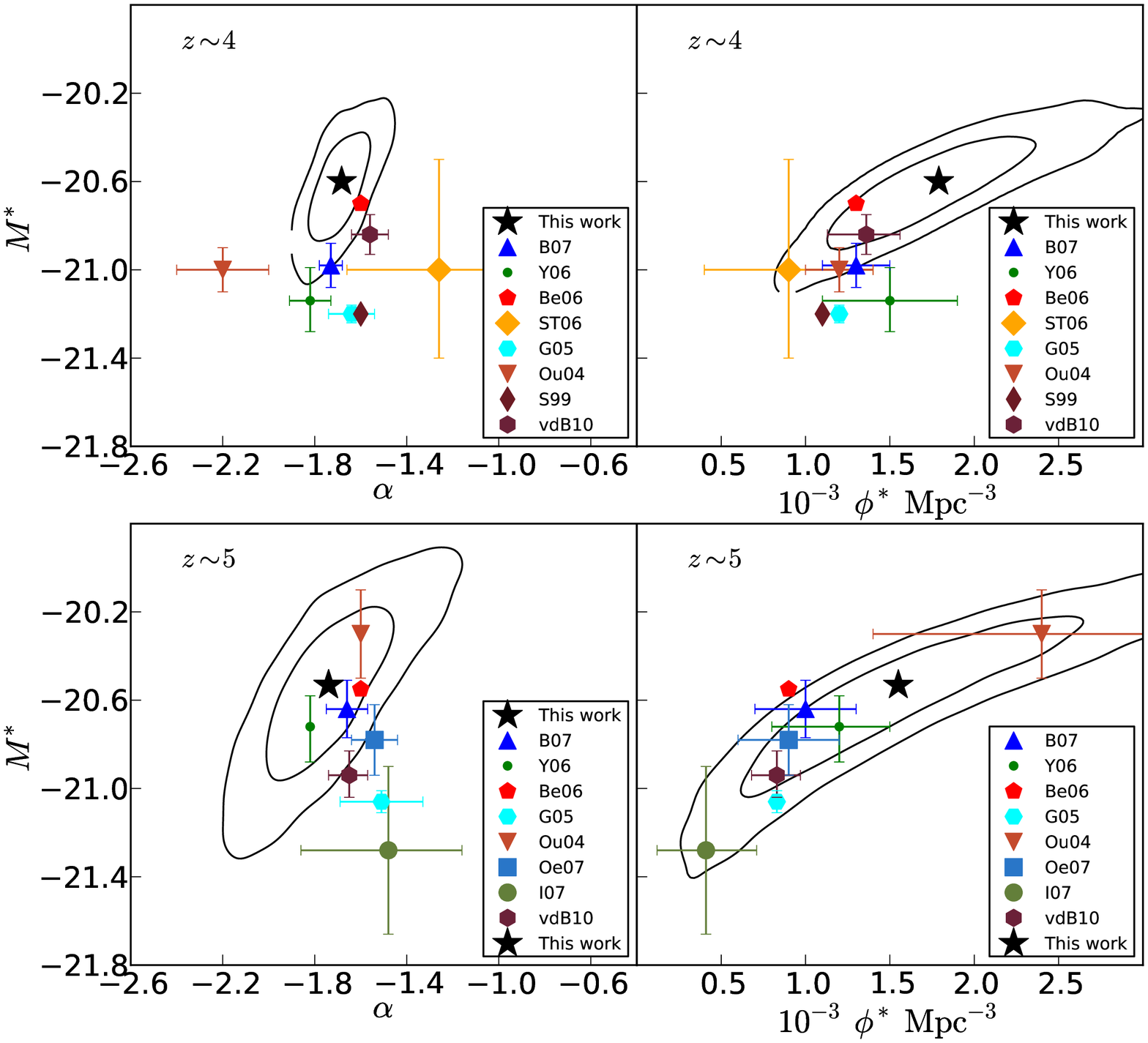}
   \figcaption{The best-fit values of $\alpha$, $M^*$, and $\phi^*$ along 
   with the 1-$\sigma$ and the 2-$\sigma$ confidence contours, of the
  Schechter function for UV LF at $z\sim 4$ (\textit{top} panel) and $z\sim 5$
  (\textit{bottom} panel) from this study and from the
  literature. The references are: \citealt{Bouwens.2007} (B07);
  \citealt{Yoshida.2006} (Y06); \citealt{Beckwith.2006} (Be06);
  \citealt{Sawicki.2006} (ST06); \citealt{Giavalisco.2005} (G05);
  \citealt{Ouchi.2004} (Ou04); \citealt{Steidel.1999} (S99); 
  \citealt{Oesch.2007} (Oe07); \citealt{Iwata.2007} (I07);
  \citealt{vanderBurg.2010} (vdB10).
  \label{fig:lfparams_z4_z5}}
\end{figure}

We compare our derived best-fit values of the Schechter function with those of other studies at similar redshifts in Figure \ref{fig:lfparams_z4_z5}. Our derived $\alpha$ and $\phi^*$ are in general consistent within 1-$\sigma$ of those from most studies; however, our $M^*$ is about $0.2-0.4$ mag fainter than previous studies. The fainter $M^*$ is possibly driven by the low completeness near the faint end, where we potentially miss many galaxies due to low surface brightness. The inclusion of photometric interlopers is likely not the origin of the discrepancy: without photometric interlopers, the best-fit $(\alpha, M^*, \phi^*)$ becomes $(-1.64, -20.48, 2.21\times10^{-3}\,\mathrm{Mpc}^{-1})$ at $z\sim 4$ and $(-1.67, -20.51, 1.70\times10^{-3}\,\mathrm{Mpc}^{-1})$ at $z\sim 5$, statistically indistinguishable from our nominal best-fit parameters. Extending our methodology to include more fields (e.g. HUDF parallel fields; \citealt{Thompson.2005,Oesch.2007}) will better constrain the LF parameters by increasing the sample size. 

However, we have reason to believe that the cosmic variance introduces very small biases in our best-fit parameter values. \citet{Trenti.2008} (see also \citealt{Robertson.2010}) has a comprehensive discussion about the effect of cosmic variance on the determination of luminosity functions at high redshift, especially in
narrow, pencil-beam fields such as HUDF. In particular, in their Section 5 they examined how the number density of high-redshift dropouts will bias the best-fit Schechter function parameters. They demonstrate that within a single field, $M^*$ tends to be fainter when the field is under-dense; however, when one combines a shallower, larger field with a deeper, smaller field, and derives the LF parameters using the unbinned maximum-likelihood method as in \citet{Sandage.1979}, biases in both $\alpha$ and $M^*$ can be reduced. Because our fitting method is the one proposed by \citet{Sandage.1979} generalized to include multiple fields, we do not expect our best-fit $\alpha$ and $M^*$ to be strongly biased due to cosmic variance.

\subsection{The Size Evolution}\label{subsec:size_evolution}
The peak of the size distribution at the same rest-frame 1500\AA\ absolute magnitude, $M_{1500}=-21$ mag, is $\Rzeroresbkpc\Rzeroresbkpcerr$ kpc for the \acsb-dropout sample (assuming $z=4$) and is $\Rzeroresvkpc\Rzeroresvkpcerr$ kpc for the \acsv-dropout sample (assuming $z=5$). If instead we calculate the peak of the size distribution at the best-fit $M^*$ of each redshift, we will find that the size distribution peaks at $1.24$ kpc at $z = 4$ (at $M^*=\mstarresb$ mag), and at $1.07$ kpc at $z = 5$ ($M^*=\mstarresv$ mag), showing slightly stronger evolution than at fixed luminosity. 

The conventional picture of \citet{Fall.Efstathiou.1980} (see also \citealt{Fall.1983,Dalcanton.1997,Mo.1998,Fall.2002}) states that the size of the gaseous disk $R_d$, formed in the center of dark-matter halo that acquired angular momentum through tidal interactions, is proportional to the virial radius $R_{\mathrm{vir}}$ of the host dark-matter halo and the spin parameter $\lambda$ of the halo: $R_d \propto \lambda R_{\mathrm{vir}}$. Assuming that the peak of the size distribution $R_0(z)$ at a given luminosity $L_0$ traces the peak of the log-normal $\lambda$ distribution, and the $\lambda$ distribution correlates weakly with redshift or halo mass (\citealt{Barnes.1987}), one comes to the conclusion that $R_0(z) \propto R_{\mathrm{vir}}(z)$. One can further relate the virial radius $R_{\mathrm{vir}}(z)$ of the halo to the circular velocity $V_c$ and the mass $M_\mathrm{H}$ of the halo through (\citealt{Ferguson.2004})
\begin{equation}\label{eq:virial_radius}
   R_{\mathrm{vir}} = \frac{V_c}{10H(z)} = \left(\frac{GM_\mathrm{H}}{100}\right)^{1/3}\ H^{-2/3}(z)
\end{equation}
so that $R_d \propto H^{-1}(z)$ if $R_{\mathrm{vir}}$ at a fixed luminosity traces constant circular velocity, or $R_d \propto H^{-2/3}(z)$ if $R_{\mathrm{vir}}$ traces constant halo mass, and $H(z)$ is the Hubble parameter. In the concordance cosmology model, the Hubble parameter as a function of $z$ is $H(z) = H_0\sqrt{\Omega_m(1+z)^3+\Omega_\Lambda}$, so at $z \geq 2$, $H^{-1}(z) \sim (1+z)^{-3/2}$ and $H^{-2/3}(z) \sim (1+z)^{-1}$. Between $z=4$ and 5, both relations predict only 20\% and 31\% changes in size, respectively. The peak size at $M_{1500}=-21$ mag evolves only by 13\% from our results, favoring only slightly the $(1+z)^{-1}$ relation. One needs data at lower redshifts to constrain the size-evolution scenario, where these two scaling relations diverge. 

Studies of size evolution of disk-like (or late-type) galaxies with similar luminosities have generally shown larger fractional changes in size at $z \leq 3$. At $z \lesssim 1.1$, using GEMS (\citealt{Rix.2004}) data, \citet{Barden.2005} found that the mean rest-frame $V$-band surface brightness of late-type galaxies with $M_V \lesssim -20$ mag brightens by about 1 mag from $z \sim 0$ to $z \sim 1$. This corresponds to a $40\%$ decrease in mean size of the galaxies with similar luminosities. \citet{Trujillo.2006}, combining FIRES (\citealt{Franx.2000}), GEMS, and SDSS (\citealt{York.2000}) data, found that at rest-frame $V$-band luminosity $L_V > 3.4 \times 10^{10}h_{70}^{-2}\,L_\odot$ (or $M_V \leq -21.53$ mag) the mean effective half-light radius of late-type galaxies is only about 30\% of that of the local SDSS sample. At high redshift, \citet{Ferguson.2004} found the mean half-light radius of galaxies selected in GOODS from $z\sim 1.4$ to $z \sim 5$ follows the $H^{-1}(z)\sim (1+z)^{-3/2}$ scaling relation. On the other hand, both \citet{Bouwens.2004} and \citet{Oesch.2010} found the mean half-light radii of LBGs, selected in HUDF and HUDF ACS-parallel fields, to roughly trace $(1+z)^{-1}$ from $z\sim 3$ to $z\sim 8$, although \citet{Oesch.2010} noted that the peak of the size distribution evolves little between $z\sim 4-8$. From the above observations of star-forming galaxies, the fractional change in size at similar luminosities is significant between $z=0-3$, but smaller at $z \gtrsim 3$. 

\begin{figure}[t]
	\epsscale{1.0}
   \plotone{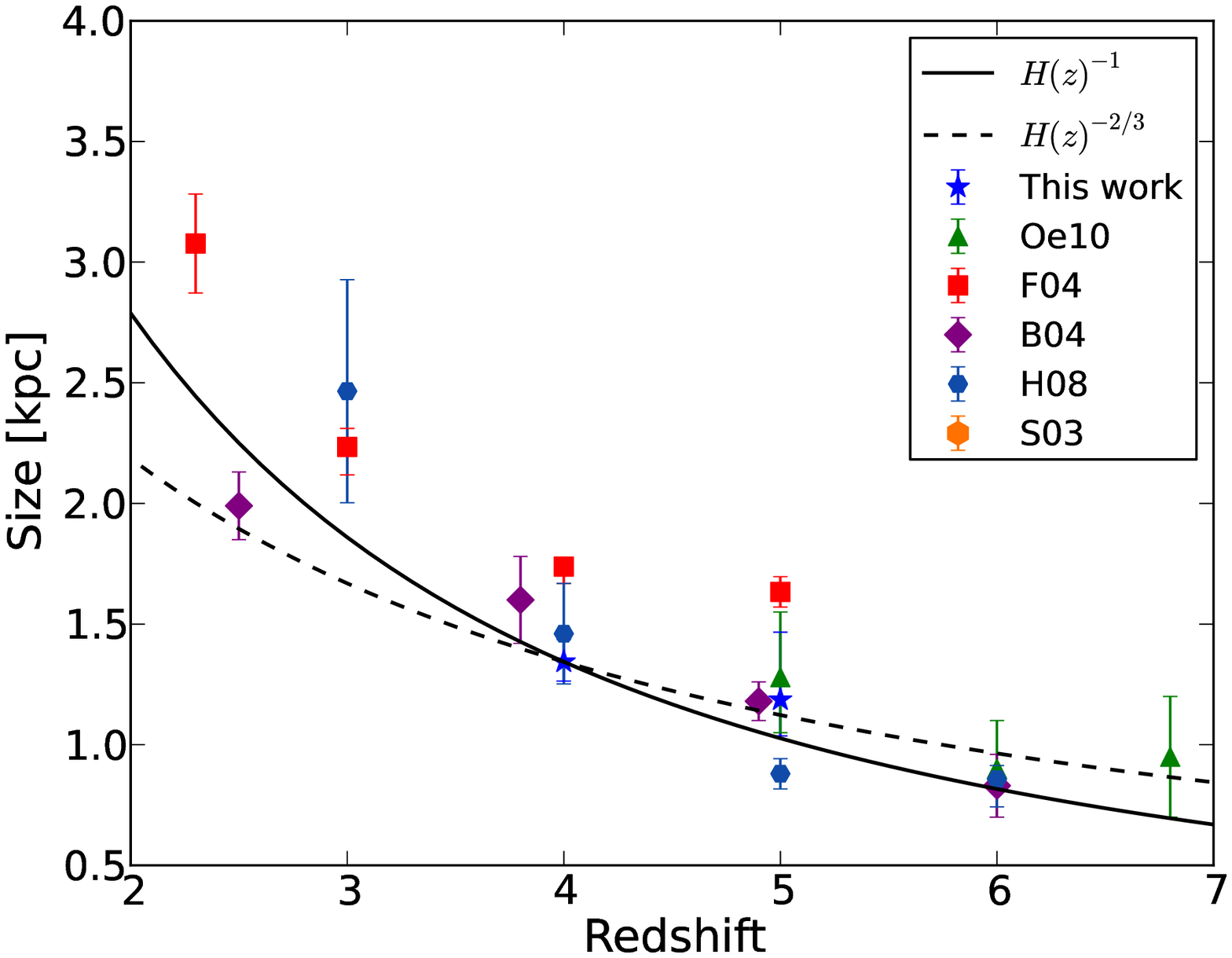}
   \figcaption{The size evolution of LBGs at $z \gtrsim 2$ from this work and previous studies. Included in this figure are results from \cite{Ferguson.2004} (F04; their mean size for LBGs between $0.7L^*_{z=3}$ and $5L^*_{z=3}$, where $L^*_{z=3}$ is the characteristic luminosity of a $z=3$ LBG from \citealt{Steidel.1999}), \cite{Bouwens.2004} (B04; the mean size for LBGs between $0.3L^*_{z=3}$ and $L^*_{z=3}$), \cite{Hathi.2008} (H08; the mean size of LBGs with spectroscopic redshifts), and \cite{Oesch.2010} (Oe10; the mean size of LBGs between $0.3L^*_{z=3}$ and $L^*_{z=3}$, similar to \citealt{Bouwens.2004}). In this paper we adopt the effective half-light radius $R_e$ that encloses half the total flux as galaxy size, therefore we try to compensate for the missing flux at the edge of a galaxy, while all other studies quoted here adopt half-light radius ($R_h$) measured by SExtractor. We explored the bias of effective half-light radius and half-light radius in Section \ref{subsec:galfitsim}. The size evolution relations $H(z)^{-1}$ and $H(z)^{-2/3}$ are plotted as the solid and dashed line, respectively, with the curves normalized to the $z=4$ size from this work. All sizes are converted to the cosmological parameters adopted in this study at the quoted mean redshift of each sample. Mean sizes at $z\lesssim 2$, at the same rest-frame wavelength as probed at higher redshifts, are required to distinguish between different evolutionary scenarios. 
   \label{fig:size_evol}}
   \end{figure}

We plot the mean sizes of LBGs at $z \gtrsim 2$ (in rest-frame UV) from a compilation of previous papers, as well as our best-fit $R_0$ at $M_{1500}=-21$ mag, as a function of redshift in Figure \ref{fig:size_evol}.  Also included in the figure are the curves $H(z)^{-1}$ and $H(z)^{-2/3}$, both normalized to the derived $R_0$ at $z=4$ of this work. Due to different measurement methods, definitions of size, corrections for measurement biases and uncertainties, and corrections for incompletenesses, systematic differences in size at the same redshift exist. (The definition of effective half-light radius that we use is also meant to enclose half the total flux from a galaxy, but it does try to compensate for the missing flux at the edge of a galaxy, unlike the half-light radius measured by SExtractor as adopted in all other papers included here.) For example, \citet{Ferguson.2004} derived the mean sizes of LBGs at $z\sim 1-5$ with rest-frame UV luminosity between $0.7L^*_{z=3}$ and $5L^*_{z=3}$, where $L^*_{z=3}$ is the characteristic luminosity of LBGs at $z=3$ from \citet{Steidel.1999}. They did not correct for sample incompleteness and measurement bias, and they reported the average size instead of the peak of the size distribution as in this paper (the mean of a log-normal distribution is larger than the peak due to the tail at the high end). Comparisons between high-$z$ LBGs and local galaxies are also hampered by different rest-frame wavelengths probed. In order to compare with the rest-frame optical sizes of local galaxies, one needs resolved images of most LBGs at $z\gtrsim 3$ in the rest-frame optical. However, currently such data are not available, and the \textit{James Webb Space Telescope} will be necessary for this line of investigation. We also wish to extend our analyses to lower-redshift samples where high-resolution rest-frame UV data are available (down to $z\sim2$; see e.g. \citealt{Hathi.2010}) to obtain consistent size measurements across a longer redshift baseline.

\subsection{Width of the Size Distribution}
We find that the width of the size distribution, $\sigma_{\ln R}$, is $\sigmaresb\sigmaresberr$ at $z \sim 4$ and $\sigmaresv\sigmaresverr$ at $z \sim 5$. This seems to indicate that the width of the size distribution at fixed luminosity is slightly larger at $z \sim 5$, and both are larger than the width of the $\lambda$ distribution ($\sigma_{\ln \lambda}\sim 0.5$) of dark matter halos found in the $N$-body simulations (\citealt{Barnes.1987,Warren.1992,Bullock.2001}).

\begin{figure}[t]
	\epsscale{1.0}
   \plotone{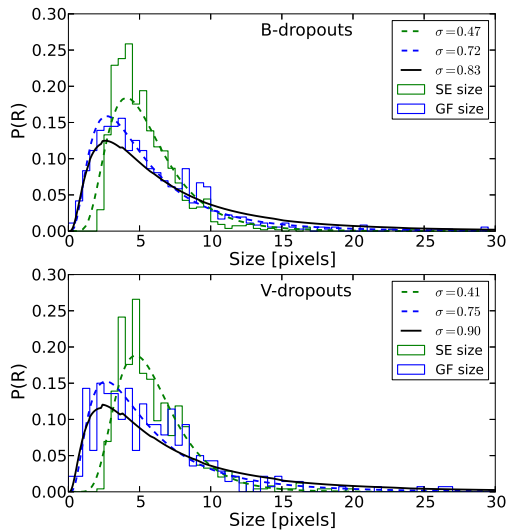}
   \figcaption{The distribution of SExtractor-measured half-light radii ($R_{50}$) and GALFIT-measured effective half-light radii ($R_e$) together with their best-fit log-normal distributions without accounting for incompletenesses or biases. Light-gray histograms and dashed curves are those for the SExtractor-measured $R_{50}$; dark-gray histograms and dashed curves are those for the GALFIT-measured $R_e$. The GALFIT-measured $R_e$ only include the ones with acceptable fits. The black solid curve is the volume-weighted sum of two log-normal distributions in GOODS and HUDF that correspond to the best-fit parameters listed in Table \ref{tab:bestfit_params}. It represents our best estimate of the underlying size distribution of the two fields. The histograms and curves are normalized to give a total probability of 1, so incompleteness is not included. The widths of the lognormal distributions $\sigma_{\ln \mathrm{R}}$ are indicated in the legends.
   \label{fig:sizedist_raw}}
\end{figure}

We plot the distribution of sizes of the sample --- both measured by GALFIT and by SExtractor --- in Figure \ref{fig:sizedist_raw}. The observed widths of size distribution, without any statistical correction for biases and uncertainties, are $\sigma_{\ln\mathrm{R}}\gtrsim 0.7$ for the GALFIT-measured effective half-light radii. In comparison, the SExtractor-measured half-light radii shows narrower distributions: the widths of the best-fit log-normal distributions are $\sigma_{\ln\mathrm{R}}\gtrsim 0.4$. Although the widths of the SExtractor-measured size distribution are more in line with the theoretical expectations ($\sigma_{\ln\lambda}\sim 0.5$), we believe that it is biased because (1) SExtractor is severely limited by the PSF size, so it is strongly biased (i.e. the measured size is bigger than the true size) when the galaxies are barely resolved, and (2) for large galaxies SExtractor is also strongly biased due to the tendency to miss the tail of the light distribution. We therefore believe that the intrinsic width of the size distribution is significantly wider than what SExtractor-measured distribution indicates, and our statistical treatment indeed shows that the intrinsic size distribution is wider than the that measured by SExtractor.

One must also consider the fact that GALFIT (or any size-measurement method) can artificially broaden the observed size distributions due to measurement uncertainties. However our results show that although GALFIT might artificially broaden the intrinsic size distributions, the size distributions after accounting for sample-selection incompleteness are even wider than the GALFIT-measured distributions. Therefore sample incompletenesses, especially at the small- and large-size ends, are more important in shaping the observed widths of the size distributions than whichever measurement method one uses.

The widths of the size distribution of local galaxies are in general smaller than what are found here for LBGs. When considering galaxies divided by their morphological types drawn from SDSS ($z \sim 0.1$), \cite{Shen.2003} found that $\sigma_{\ln \mathrm{R}}\sim 0.3$ at the bright-end, while the width increases to $\sigma_{\ln \mathrm{R}}\sim 0.5$ at the faint-end. This is true when considering either late-type galaxies alone or early-type galaxies alone. Furthermore, both \cite{deJong.2000} and \cite{Courteau.2007} derived that $\sigma_{\ln \mathrm{R}}\sim 0.3$ for their local spiral galaxies. All of the above results show that the size distribution for local disk galaxies is narrower than the $\lambda$ distribution of DM halos, and mechanisms to systematically remove disk galaxies residing in small-$\lambda$ and large-$\lambda$ halos such as bar instability and supernova feedback are suggested. Given the wide size distribution derived for LBGs in this work, it is possible that LBGs sample the whole spectrum of $\lambda$, and some aspects of baryon physics further broaden the size distribution. In addition, there are also hints of wider size distributions than that of late-types or early-types alone when we consider both types together (e.g. \citealt{Fall.1983,Romanowsky.2012}). Because it is still unknown whether the majority of LBGs are rotation-supported disks, the relation between the width of the size distributions of LBGs and that of $\lambda$ is still not clear.

\subsection{The Size-Luminosity Relation}\label{subsec:sizelum_relation}
\begin{figure}[t]
   \plotone{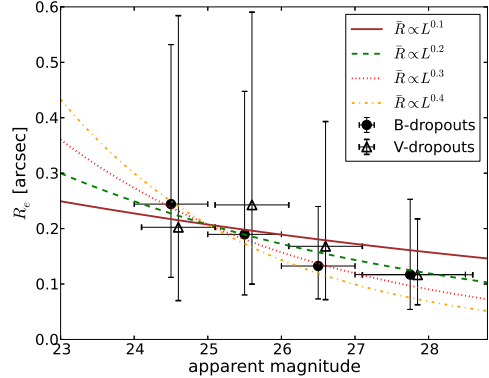}
   \figcaption{The mean of the log-normal size distribution in 
   apparent magnitude bins (24, 25), (25, 26), (26, 27), and (27, 28.5) for
   both \acsb-dropouts and \acsv-dropouts. The apparent magnitude is in 
   the \acsi-band for the \acsb-dropouts and in the \acsz-band for the \acsv-
   dropouts. The error bars in the $y$-direction are the $\pm 1\sigma$ width of the log-normal
   distribution; the error bars in the $x$-direction are the bin width in magnitude. 
   The size distributions in each magnitude bin are fitted when $\alpha$, $M^*$,
   and $\beta$ are held fixed to the best-fit values in Table \ref{tab:bestfit_params}. 
   Size-luminosity relations of the form $\bar{R}
   \propto L^\beta$ with different $\beta$ are also plotted. \label{fig:sizemag}}
\end{figure}

Adopting the power-law form of the size-luminosity relation $\bar{R} \propto L^\beta$, we find that $\beta = \betaresb\betaresberr $ for the \acsb-dropouts and $\beta = \betaresv\betaresverr$ for the \acsv-dropouts. The trend between the peak of the size distribution $\bar{R}(L)$ and luminosity $L$, without accounting for incompletenesses at the low-surface brightness end, is steeper due to selection biases against low-surface brightness galaxies. The observed $\beta$, without accounting for incompleteness, is 0.32 for the \acsb-dropouts and 0.30 for the \acsv-dropouts. We also plot in Figure \ref{fig:sizemag} the peak of the log-normal size distribution $\bar{R}(L)$ and the width $\sigma_{\ln R}$ in different apparent magnitude bins, when $\alpha$, $M^*$, and $\beta$ are held fixed to the best-fit values. Different scaling relations between the mean size $\bar{R}$ and luminosity $L$ are also plotted. After accounting for incompletenesses, the size-luminosity relation is still inconsistent with being flat.

\begin{figure}[t]
   \plotone{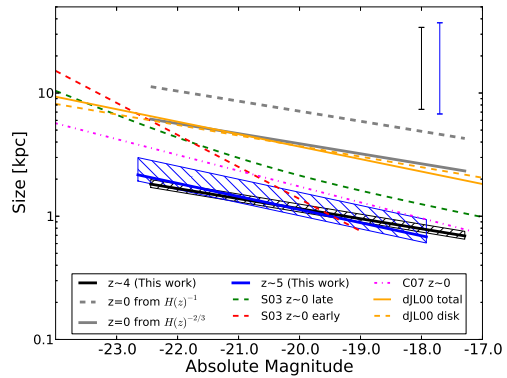}
   \figcaption{The best-fit rest-frame 1500 \AA\ size-luminosity ($RL$) relations of $z\sim 4$ and $z\sim 5$ LBGs (black and blue lines, respectively), and the hatched area shows the uncertainties of the normalization $R_0$. Also included are the best-fit size-luminosity relations for local galaxies from the literature. The slopes of the size-luminosity relations are listed in Table \ref{tab:RL_reln}. Note that the local size-luminosity relations are in rest-frame optical wavelengths instead of the UV wavelengths probed for high-$z$ LBGs. The green and red dashed lines are the SDSS $r$-band late-type and early-type galaxies from \cite{Shen.2003} (S03) with $M_r$ between $-24$ and $-16$ mag for early-types, and $-24$ and $-19$ mag for late-types. The magenta dot-dashed line is from \cite{Courteau.2007} (C07); the $I$-band luminosity $L_I$ of their local disk galaxies range from $10^{8.5}$ to $10^{11.5}$ $L_{I,\odot}$. The orange solid and dashed lines are for the disk+bulge and disk-only components of the galaxies from \cite{deJong.2000} (dJL00); the $I$-band absolute magnitude $M_I$ of their disk galaxy range from $-25.5$ to $-16.5$ mag. The slopes of LBGs and local disk galaxies are similar, but the offset in normalization reveals the size evolution between $z\sim 4-5$ and $z\sim 0$. The error bars on the top-right corner show the width of the size distribution for LBGs. The projected size-luminosity relation from the best-fit relation at $z\sim 4$ are plotted with a gray solid line (following $H(z)^{-1}$, which is $\sim (1+z)^{1.5}$ at $z>2$) and a gray dashed line (following $H(z)^{-2/3}$, which is $\sim (1+z)^{-1}$ at $z>2$). It is interesting that the prediction from $H(z)^{-2/3}$ matches the \citet{deJong.2000} results very well. \label{fig:RL_reln}}
\end{figure}

The observed size-luminosity ($RL$) relation of low-$z$ disk galaxies in the optical bands has been shown to have similar power-law index $\beta$ as our result for the high-$z$ galaxies. We compile the size-luminosity relations for local galaxies in Table \ref{tab:RL_reln}. For example, \citet{deJong.2000} observed that $R_d \propto L^{0.25}$ for their local disk galaxy sample with types later than Sb. \citet{Shen.2003} observed that $R_e \propto L^{0.26}$ in the faint-end of their late-type galaxy sample (defined as $n < 2.5$) drawn from SDSS, where $R_e$ is their best-fit \sersic\ effective half-light radius, although they also observed that at the bright end of their late-type galaxies the observed slope is significantly steeper ($\beta \sim 0.5$). \citet{Courteau.2007} observed that $R_d \propto L^{0.32}$ for their local disk galaxy sample, including both field and cluster galaxies, and also observed a type-dependent trend in $\beta$ such that later-type disk galaxies have shallower slope ($\beta = 0.25$ for their Sd-type sample). On the other hand, the observed $RL$ relation for local early-type galaxies generally have steeper slopes: $\beta \sim 0.5-0.6$ in general (\citealt{Shen.2003}). We plot the $RL$ relations from local galaxies along with the derived $RL$ relations for out LBGs in Figure \ref{fig:RL_reln}. The agreement between the slopes of the local $RL$ relations and those for the LBGs in this study is interesting. Although most galaxies in our LBG sample have $n < 2.5$, which will be classified as late-type galaxies based on \sersic\ index only, whether they are disk galaxies is still uncertain. For example, LBGs have ellipticity distributions inconsistent with being drawn from an optically-thin disk (\citealt{Ravindranath.2006,Law.2012}). On the other hand, the normalization of the $RL$ relation between different redshifts reveals the amount of size evolution between these two epochs. The similarities between the size-luminosity relation of the local late-type galaxies and high-$z$ LBGs suggests that their formation histories share some similarities, or even that LBGs are the progenitors of local disk galaxies. Furthermore, if one scales the best-fit $RL$ relation at $z\sim 4$ according to $H(z)^{-2/3}$, the resultant $RL$ relation matches the local relations very well (the dark-gray solid line in Figure \ref{fig:RL_reln}).

\citet{deJong.2000} (see also \citealt{Fall.1983,Dalcanton.1997,Mo.1998}) offered a derivation of the expected value of $\beta$ for disk galaxies. We saw in Section \ref{subsec:size_evolution} that in the standard picture of disk formation, $R_\mathrm{d} \propto \lambda R_{\mathrm{vir}}$, and the virial radius $R_{\mathrm{vir}}$ is proportional to the circular velocity $V_c$ at a given redshift (Eq. \ref{eq:virial_radius}). The Tully-Fisher relation (\citealt{Tully.1977}) of local disk galaxies in the $i$-band shows that $V_c \propto L^\beta$, where $\beta \sim 0.29$ (\citealt{Courteau.2007}). This translates to $R_d \propto \lambda V_c \propto \lambda L^\beta$. The best-fit $\beta$ of our sample, \betaresb\ for the \acsb-dropouts and \betaresv\ for the \acsv-dropouts, are slightly smaller than the prediction by the simple argument, but are still consistent with the TF slope of local disk galaxies.

\section{Summary}\label{sec:summary}

We study the bivariate size-luminosity distributions of Lyman Break galaxies (LBGs) at redshifts around 4 and 5. We select LBGs from the GOODS and the HUDF fields to constrain the size-luminosity distribution. We break down the possible systematic errors into two components --- one from dropout selection, the other from size and magnitude measurement by GALFIT. We then perform comprehensive simulations to quantify these errors, and we develop numerical tools to apply the corrections statistically to our model --- both in magnitude and size bins. The bivariate size-luminosity distribution model is a combination of Schechter function and a log-normal distribution, and our best-fit parameters, after all the corrections, are summarized in Table \ref{tab:bestfit_params}.

The best-fit parameters for the luminosity function show a slightly shallower faint-end slope $\alpha$, slightly brighter characteristic absolute magnitude $M^*$, and higher normalization constant $\phi^*$ of the luminosity function at $z\sim4$ than at $z\sim 5$. However the parameters are consistent with no evolution within 1-$\sigma$. The parameters for size distribution also show mild evolution between $z\sim 4$ and 5, with \acsb-dropouts having larger peak size $R_0$ at $M_{1500}=-21$ mag, narrower width $\sigma_{\ln R_e}$ of size distribution, and shallower size-luminosity relation slope $\beta$. The faint-end slope $\alpha$ and the slope of size-luminosity relation $\beta$ show a mild correlation, suggesting the importance of size-luminosity relation in the completeness correction of the faint end of luminosity function.

A significant result from our derived size distribution is that the widths of the size distributions for LBGs at $z\sim4$ and 5 are larger than the width of the spin parameter distribution observed in cosmological $N$-body simulations, as opposed to local disk galaxies which have narrower size distributions. We argue that the narrow size distributions observed could be due to the bias inherent in the size-measurement method. This also implies that the LBGs sample the full range of the $\lambda$ distribution. The slopes of the size-luminosity relations $\beta$ are consistent with those for local late-type galaxies, but are shallower than local early-type galaxies. Given the uncertain disk nature of LBGs at these redshifts, the detailed connection between size and $\lambda$ is not clear.

We plan to extend our study to include other deep ACS fields (e.g. HUDF05, \citealt{Oesch.2007}) in future work to increase the sample size at the redshifts probed in this work. Furthermore, we will also extend our investigations to lower redshifts to lengthen our redshift baseline. Size distributions from $z\lesssim 3$ are also crucial in determining the functional form of size evolution as a function of redshift. However we are limited to the rest-frame UV for measuring the size distributions of LBGs at $z\gtrsim 5$; high-resolution, rest-frame optical data for LBGs are not available until \textit{JWST} comes online. With \textit{JWST}, it will be possible to compare rest-frame optical sizes of local and high-redshift galaxies and eliminate the uncertainties introduced by morphological $K$-corrections.

\begin{deluxetable*}{cccc}
\tablecaption{The Slope of the Size-Luminosity Relations and the Width of the Log-normal Size Distribution of Local Galaxies and LBGs at $z\sim 4$ and 5. \label{tab:RL_reln}}
\tablehead{\colhead{Sample} & \colhead{Reference} & \colhead{Slope ($\beta$)\tablenotemark{a}} & \colhead{Width ($\sigma_{\ln R}$)}}
\startdata
$z\sim 4$ LBG & This work & $\betaresb\betaresberr$ & $\sigmaresb\sigmaresberr$ \\
$z\sim 5$ LBG & This work & $\betaresv\betaresverr$ & $\sigmaresv\sigmaresverr$ \\
$z\sim 0$ late-type\tablenotemark{b} & \citealt{Shen.2003}\tablenotemark{c} &  $0.26$\tablenotemark{h}  &  $0.45$ \\
$z\sim 0$ late-type\tablenotemark{d} & \citealt{Shen.2003} & $0.51$ & $0.27$ \\
$z\sim 0$ early-type\tablenotemark{e} & \citealt{Shen.2003} & $0.65$ & $0.45$/$0.27$\\
$z\sim 0$ disk & \citealt{Courteau.2007}\tablenotemark{f} & $0.321\pm 0.010$ & $0.325$\\
$z\sim 0$ disk+bulge & \citealt{deJong.2000}\tablenotemark{g} & $0.253\pm 0.020$ & $0.28\pm 0.02$\\
$z\sim 0$ disk only & \citealt{deJong.2000} & $0.214\pm 0.025$ & $0.36\pm 0.03$\\

\enddata
\tablenotetext{a}{Assuming $R \propto L^{-\beta}.$}
\tablenotetext{b}{Galaxies with $n<2.5$ and $r$-band absolute magnitude $M_r \geq -20.91$ mag.}
\tablenotetext{c}{The quoted size is the $r$-band \sersic\ effective half-light radius.}
\tablenotetext{d}{Galaxies with $n<2.5$ and $r$-band absolute magnitude $M_r \leq -20.91$ mag.}
\tablenotetext{e}{Galaxies with $n>2.5$; the widths are the same as late-type galaxies in both the faint end and the bright end.}
\tablenotetext{f}{The quoted size is the $I$-band exponential scale length corrected for inclination.}
\tablenotetext{g}{The quoted size is the $I$-band half-light radius of an exponential light profile extrapolated to infinity.}
\tablenotetext{h}{\cite{Shen.2003} did not give formal uncertainties of the derived slopes and scatter of their size-luminosity relations; \cite{Courteau.2007} did not give their formal uncertainty of the derived scatter.}

\end{deluxetable*}

We are grateful to the referee for the comments that lead to significant improvements of this paper. We thank Michael Fall, Steve Finkelstein, Nimish Hathi, Mauro Giavalisco, Russell Ryan, Massimo Stiavelli, and Sara Salimbeni for useful discussions and help with the simulations. We also thank Chien Y. Peng for producing GALFIT. This work is supported by program number HST 60-12060, which was provided by NASA through a grant from the Space Telescope Science Institute, which is operated by the Association of Universities for Research in Astronomy, Inc., under NASA contract NAS5-26555.

\bibliographystyle{apj}
\bibliography{bivariate}
\appendix
\section{Correcting For Dropout Selection and GALFIT Measurement}
\subsection{Applying Dropout-Selection Function $P(M, R, z)$}\label{appendix:selfunc}
Here we elaborate on the details of applying the dropout-selection function $P(M, R_e, z)$ on the size-luminosity distribution (Eq. \ref{eq:bivariate}) on a grid of $(M, \log_{10} R_e)$. We also explain how we apply the dropout-selection function in the two fields separately, and how we combine the two fields to calculate the total likelihood.

For clarity first we define several variables:
\begin{itemize}
  \item $M$: the \textit{intrinsic} 1500\ \AA\ absolute magnitude
  \item $m$ and $R$: the \textit{intrinsic} apparent magnitude and effective half-light radius (we will omit the subscript $e$ in $R_e$ for clarity of notations)
\end{itemize}

Also the superscripts G stands for GOODS and H stands for HUDF in the equations that follow.

In the GOODS-depth images, the kernel can be expressed as the selection function $P^{\mathrm{G}}(M, \log_{10} R, z)$, defined as the fraction of sources with input $M$, $\log_{10} R$, and $z$ that is detected \textit{and} selected as a dropout by the color criteria. In this study we calculate the selection function $P^{\mathrm{G}}(M, \log_{10} R, z)$ in bins of $(M, \log_{10} R, z)$, such that $P^{\mathrm{G}}_{ijk}$ is the completeness with absolute magnitude in the range $(M_i, M_i + \Delta M)$, size in the range $(\log_{10} R_j, \log_{10} R_j + \Delta \log_{10} R)$, and redshift in the range $(z_k, z_k + \Delta z)$. The bin size $(\Delta M, \Delta \log_{10} R, \Delta z)$ is selected to be (0.5, 0.2, 0.1), and we implicitly assume that the selection function stays the same within the bin. Usually the dropout selection function $P^{\mathrm{G}}_{ijk}$ is only greater than zero within $\Delta z \sim 0.5$ around the central redshift ($z=4$ for \acsb-dropouts and $z=5$ for \acsv-dropouts).

The first step towards building the expected bivariate distribution of observed LBGs is to apply the dropout selection function $P^{\mathrm{G}}_{ijk}$ to the theoretical distribution (Eq. \ref{eq:bivariate}). We define $\Psi$, \textit{sans} the normalization factor $\phi^*$, on a grid of $(M_p, \log_{10} R_q)$ where $p$ and $q$ are the indices of the ``pixels''. We choose the pixel size to be $(dM, d\log_{10} R) = (0.02, 0.02)$. We apply the dropout-selection function in the following way: for each kernel within the range $M_i\leq M < M_i+\Delta M$ and $\log_{10} R_j \leq \log_{10} R < \log_{10} R_j + \Delta \log_{10} R$, we calculate the contribution to the number density of LBGs predicted by $\Psi$ in this bin, weighted by the volume in each redshift interval and the dropout selection function $P^{\mathrm{G}}_{ijk}$. In GOODS,
\begin{align}
   \theta^{\mathrm{G}}_{ij}(m_r, \log_{10} R_q) &= \sum_{p,k} \Psi_{ij}(M_p, \log_{10} R_q)\,P^{\mathrm{G}}_{ijk}\,
      \frac{dV}{dz_k}\,\delta(M_p - \mathrm{DM}(z_k) - K - m_r)\label{eq:theta1ij}\\
   &\equiv \sum_{ij} \Psi_{ij} \circledast P^{\mathrm{G}}_{ijk}\label{eq:theta1ijstar}
\end{align}   
where
\begin{align}
   \Psi_{ij}(M_p,\log_{10} R_q) &\equiv 
      \begin{cases}
         \Psi(M_p,\log_{10} R_q) &\mathrm{if}\ M_i\leq M_p < M_i+\Delta M\ \mathrm{and}\\
                                   & \log_{10} R_j \leq \log_{10} R_q < \log_{10} R_j + \Delta \log_{10} R\\
         0 &\mathrm{otherwise}
      \end{cases}
\end{align}
and $dV/dz_k$ is the volume element within the surveyed area and the redshift range $(z_k, z_k+\Delta z)$. The Dirac delta function $\delta(M_p - \mathrm{DM}(z_k) - K - m_r)$ facilitates the conversion from absolute magnitude $M$ to apparent magnitude $m$ so that the result is a distribution $\theta^{\mathrm{G}}_{ij}$ defined on the grid of $(m, \log_{10} R)$ with the same pixel size $(dm, d\log_{10} R)=(0.02,0.02)$. In the argument of the delta function, $\mathrm{DM}(z_k)$ is the distance modulus to redshift $z_k$ (calculated using the assumed cosmological parameters), and $K$ is the $k$-correction term that converts from rest-frame 1500 \AA\ magnitude to the apparent magnitude in the chosen band. Since we choose to use the bandpasses that are closest to rest-frame 1500 \AA, the $k$-correction term is in general very small. Thus Eq. \ref{eq:theta1ij} is the contribution to the distribution of $(m, \log_{10} R)$ from the model $\Psi$ within $M_i\leq M < M_i+\Delta M$ and $\log_{10} R_j \leq\log_{10} R\leq \log_{10} R_j + \Delta \log_{10} R$. Lastly, Eq. \ref{eq:theta1ijstar} is a shorthand notation for the application of the dropout-selection function to the model $\Psi$. 

The resultant distribution of number density, combining $\Psi$ in all absolute-magnitude and effective-radius bins and after applying the dropout selection function, is 
\begin{equation}\label{eq:theta1}
   \theta^{\mathrm{G}}(m_r, \log_{10} R_q) = \sum_{i,j} \theta^{\mathrm{G}}_{ij}(m_r, \log_{10} R_q)
\end{equation}

Similarly, we construct the distribution after applying the dropout-selection in the HUDF using the same procedure:
\begin{align}
   \theta^{\mathrm{H}}_{ij}(m_r, \log_{10} R_q) &= \sum_{p,k} \Psi_{ij}(M_p, \log_{10} R_q)\,
      P^{\mathrm{H}}_{ijk}\,\frac{dV}{dz_k}\,
      \delta(M_p - \mathrm{DM}(z_k) - K - m_r)\label{eq:theta2ij}\\
   &\equiv \Psi_{ij} \circledast P^{\mathrm{HUDF}}_{ijk}\label{eq:theta2ijstar}\\
   \theta^{\mathrm{H}}(m_r, \log_{10} R_q) &= \sum_{i,j} \theta^{\mathrm{H}}_{ij}(m_r, \log_{10} R_q)
\end{align}

\subsection{Applying GALFIT Transfer Functions}\label{appendix:gfkernel}
The second set of transfer-function (TF) kernels quantifies the GALFIT-measurement bias, scatter, and completeness of GALFIT giving reasonably good fits. It relates the \textit{intrinsic} apparent magnitude and effective half-light radius to the \textit{observed} apparent magnitude and effective half-light radius. Therefore we define
\begin{itemize}
  \item $m'$, $R'$: the \textit{measured} apparent magnitude and effective half-light radius
  \item $\delta m \equiv m_{\mathrm{measured}} - m_{\mathrm{intrinsic}} = m' - m$
  \item $\delta \log_{10} R \equiv \log_{10} R_{\mathrm{measured}} - \log_{10} R_{\mathrm{intrinsic}} = \log_{10} R' - \log_{10} R$
\end{itemize}

Again the superscripts G stands for GOODS and H stands for HUDF in the equations that follow.

Similar to the case of dropout selection TF kernels, we use a bin size of 0.5 in $m$ and 0.2 in $\log_{10} R$ and calculate GALFIT TF kernels in each intrinsic magnitude and size bin, assuming that the kernel property is uniform within the bin. The GALFIT TF kernels in each bin are probability density functions (PDF) defined on a pixel grid of $(\delta m, \delta \log_{10} R)$ within pixel width $(dm, d\log_{10} R) = (0.02, 0.02)$. The value of the GALFIT TF kernel at $(\delta m, \delta \log_{10} R)$ is the probability that the measured magnitude of the object is off by an amount of $\delta m$, and the size is off by an amount $\delta \log_{10} R$ due to measurement. In the $i'$-th apparent magnitude bin and the $j'$-th size bin, in which $m_{i'} \leq m < m_{i'} + \Delta m$ and $\log_{10} R_{j'}\leq \log_{10} R < \log_{10} R_{j'}+ \Delta \log_{10} R$, the GALFIT TF kernel in GOODS is denoted as $T^{\GF,\mathrm{G}}_{i'j'}(\delta m, \delta \log_{10} R)$.

Since each kernel $T^{\GF,\mathrm{G}}_{i'j'}(\delta m, \delta \log_{10} R)$ only applies in its own bin, we calculate the contribution to the \textbf{expected distribution of observed magnitude and size} in each bin as
\begin{subequations}
   \begin{align}
      \psi^{\mathrm{G}}_{i'j'}(m'_s,\log_{10} R'_l) =&
\sum_{q,r}\theta^{\mathrm{G}}_{i'j'}(m_r,\log_{10} R_q)\,T^{\GF,\mathrm{G}}_{i'j'}(m'_s-m_r, \log_{10} R'_l-\log_{10} R_q)\\
      \equiv &\ \theta^{\mathrm{G}}_{i'j'} \ast T^{\GF,\mathrm{G}}_{i'j'}\label{eq:psi1conv}
   \end{align}
\end{subequations}
where 
\begin{equation}\label{theta1i'j'}
   \theta^{\mathrm{G}}_{i'j'}(m_r,\log_{10} R_q) \equiv 
   \begin{cases}
      \theta^{\mathrm{G}}(m_r,\log_{10} R_q) & \mathrm{if}\ m_{i'}\leq m _r< m_{i'}+\Delta m\ \mathrm{and}\\
                                       & \log_{10} R_{j'} \leq \log_{10} R _q<
\log_{10} R_{j'}+\Delta \log_{10} R\\
      0 & \mathrm{otherwise}
   \end{cases}
\end{equation}
The distribution of $(m',\log_{10} R')$, after applying the GALFIT transfer
functions in all bins, is the sum
of $\psi^{\mathrm{G}}_{i'j'}$:
\begin{equation}\label{eq:theta1}
   \psi^{\mathrm{G}}(m'_s, \log_{10} R'_l) = \sum_{i',j'} \psi^{\mathrm{G}}_{i'j'}(m'_s, \log_{10} R'_l)
\end{equation}

Similarly in the HUDF, for the same dropout sample, one can calculate the expected distribution of 
$(m', \log_{10} R')$ as
\begin{subequations}
   \begin{align}
      \psi^{\mathrm{H}}_{i'j'}(m'_s,\log_{10} R'_l) =& \sum_{q,r}\theta^{\mathrm{H}}_{i'j'}(m_r,\log_{10} R_q)\,T^{\GF,\mathrm{H}}_{i'j'}(m'_s-m_r, \log_{10} R'_l-\log_{10} R_q)\\
      &\equiv \ \theta^{\mathrm{H}}_{i'j'} \ast T^{\GF,\mathrm{H}}_{i'j'}\label{eq:psi2conv}
   \end{align}
\end{subequations}
\begin{equation}\label{eq:theta2}
   \psi^{\mathrm{H}}(m'_s, \log_{10} R'_l) = \sum_{i',j'} \psi^{\mathrm{H}}_{i'j'}(m'_s, \log_{10} R'_l)
\end{equation}

Eqs \ref{eq:psi1conv} and \ref{eq:psi2conv} say that the application of TF kernels
is equivalent to convolution of $\theta$ and $T^{GF}$ on a discrete grid
in each bin. And Eqs 
\ref{eq:theta1} and \ref{eq:theta2} say that the final distribution is the sum of the 
contributions from all $(m', \log_{10} R')$ bins.

\subsection{Estimated Interloper Contribution}\label{subsec:interloper_contrib}
We estimate the contribution from interlopers due to photometric scatter by combining the estimated interloper fraction as a function of apparent magnitude and size in Figure \ref{fig:contam_frac}. We do not know exactly the size distribution of the interlopers, so we use the size distribution of \textit{all} sources in the field as an approximation. In the GOODS fields, if the size distribution of all non-dropout sources is $\mathcal{S}^{\mathrm{G}}(R'_l)$, and the interloper number density as a function of apparent magnitude is $\mathcal{I}^{\mathrm{GOODS}}(m'_s)$, then the interloper contribution is
\begin{equation}
   I^{\mathrm{G}}(m'_s, R'_l) = \mathcal{I}^{\mathrm{G}}(m'_s)\,\mathcal{S}^{\mathrm{G}}(R'_l)
\end{equation}
Similarly, in HUDF
\begin{equation}
   I^{\mathrm{H}}(m'_s, R'_l) = \mathcal{I}^{\mathrm{H}}(m'_s)\,\mathcal{S}^{\mathrm{H}}(R'_l)
\end{equation}

\subsection{The Resultant Bivariate Size-Luminosity Distribution}
After accounting for dropout-selection and
GALFIT-measurement, the expected distributions on a discrete pixel grid 
in the GOODS and HUDF are constructed as
\begin{equation}
   \begin{aligned}
      \psi^{\mathrm{G}}(m'_s,\,\log_{10} R'_l) &= \sum_{i',j'} \left[ \sum_{i,j} \Psi_{ij} 
         \circledast P^{\mathrm{G}}_{ijk} \right] \ast T^{\GF,\mathrm{G}}_{i'j'} \\
      \psi^{\mathrm{H}}(m'_s,\,\log_{10} R'_l) &= \sum_{i',j'} \left[ \sum_{i,j} \Psi_{ij} 
         \circledast P^{\mathrm{H}}_{ijk} \right] \ast T^{\GF,\mathrm{H}}_{i'j'}
   \end{aligned}
\end{equation}
We add the interloper contributions (Section \ref{subsec:interloper_contrib}) $I^{\mathrm{G}}$
and $I^{\mathrm{H}}$ to $\psi^{\mathrm{G}}$ and $\psi^
{\mathrm{H}}$ to compare with the 
measured magnitudes and sizes of our LBG sample and derive the best-fit parameters.

\end{document}